# The Matter of Chance: Auditing Web Search Results Related to the 2020 U.S. Presidential Primary Elections Across Six Search Engines

Aleksandra Urman, Mykola Makhortykh, and Roberto Ulloa

**Abstract:** We examine how six search engines filter and rank information in relation to the queries on the U.S. 2020 presidential primary elections under the default — that is nonpersonalized — conditions. For that, we utilize an algorithmic auditing methodology that uses virtual agents to conduct large-scale analysis of algorithmic information curation in a controlled environment. Specifically, we look at the text search results for "us elections," "donald trump," "joe biden," "bernie sanders" queries on Google, Baidu, Bing, DuckDuckGo, Yahoo, and Yandex, during the 2020 primaries. Our findings indicate substantial differences in the search results between search engines and multiple discrepancies within the results generated for different agents using the same search engine. It highlights that whether users see certain information is decided by chance due to the inherent randomization of search results. We also find that some search engines prioritize different categories of information sources with respect to specific candidates. These observations demonstrate that algorithmic curation of political information can create information inequalities between the search engine users even under nonpersonalized conditions. Such inequalities are particularly troubling considering that search results are highly trusted by the public and can shift the opinions of undecided voters as demonstrated by previous research.

Search engines play a crucial role in today's high-choice media environment (Van Aelst et al., 2017). The rapid growth of the volume of available information dictates the need for large-scale filtering and ranking of information sources. Without automated mechanisms for prioritizing reliable and relevant sources, users would be overwhelmed by the abundance of information. Thus, search engines turn into major information gatekeepers (Laidlaw, 2010; Schulz et al., 2005; Wallace, 2018) with their ranking and filtering mechanisms directing the information that is received by the users. By doing so, these mechanisms of algorithmic curation can influence users' beliefs and decisions and, in some cases, also reinforce their existing biases (Noble, 2018; White & Horvitz, 2015).

The leverage of search engine algorithms on information filtering and ranking is of particular concern in the context of political campaigns. Previous research has shown that merely the way the results are ranked can shift the voting preferences of undecided voters by 20% or more with the potential shift being as high as 80% for some demographic groups (Epstein & Robertson, 2015). While the effect size depends on the share and demographics of undecided voters as well as the level of Internet penetration in the country, it highlights the potential influence that search engines have over election outcomes. Hence, search engine corporations are increasingly called to take responsibility for preventing biases in search results and providing citizens with consistent and reliable information (Elgesem, 2008; Hinman, 2008).

The present study builds on the previous research on diversity, biases, and discrepancies in political web search results (i.e., Diakopoulos et al., 2018; Haim et al., 2018; Puschmann, 2019; Robertson, Jiang, et al., 2018; Steiner et al., 2020). Using an algorithmic auditing methodology (Mittelstadt, 2016) that builds on that proposed by Haim and colleagues (Haim et al., 2017; Haim, 2020), we investigate the curation of information on search engines under the default — that is, nonpersonalized — filtering and ranking conditions. Our methodological approach helps to further advance the field of algorithmic impact auditing as it allows tracing the effects of randomization of search results at scale. In-built randomization is a factor that can lead to major differences in search output (Makhortykh et al., 2020), yet it has been largely overlooked in the previous search engine auditing studies (i.e., Haim et al., 2018; Hannak et al., 2013; Puschmann, 2019; Robertson, Jiang, et al., 2018; Robertson, Lazer, et al., 2018; Trielli & Diakopoulos, 2019). The details on the methodology are outlined in the relevant section.

We contribute to the existing research on the representation of political topics in search results by examining how search engines distribute information about the candidates for the U.S. 2020 presidential elections during the primaries using a set of the following search queries: "us elections," "joe biden," "donald trump," and "bernie sanders." We compare our observations to those of the previous studies conducted in the context of the 2016 U.S. presidential elections (Kulshrestha et al., 2019; Trielli & Diakopoulos, 2019) and discuss potential implications of our observations.

In the current study, we explore the differences in the search results provided by six major search engines (Google, Yahoo!, Bing, DuckDuckGo, Baidu, and Yandex) for the queries related to the 2020 U.S. presidential elections during the early stage of the election campaign. Specifically, we scrutinize the results mentioning the U.S. elections, incumbent president Donald Trump, Bernie Sanders, and Joe Biden. We included queries on both, Sanders and Biden, as they were the two major contenders for the Democratic presidential nomination at the time when the experiment was conducted—1 week before the so-called Super Tuesday (the day when the biggest number of the U.S. states hold primaries).

In the context of politics, search engines are of utmost importance since, at least in the Western democracies, they are the first place where people look for political information (Dutton et al., 2017). At the same time, the general public tends to highly trust web search output (Pan et al., 2007; Schultheiß et al., 2018) despite the fact that several studies have shown that search results can exhibit racial and gender biases (Kay et al., 2015; Noble, 2018; Singh et al., 2020). Because the way search results are ranked can shift political opinions of undecided voters (Epstein & Robertson, 2015), it is important to investigate how political information is curated by search engines. To do so, we examine the differences in search outputs for the queries related to the 2020 U.S. presidential elections and aim to answer the following research questions:

> **Research Question 1**: How large are the differences in the results provided by various search engines under the default (i.e., nonpersonalized) selection and prioritization of information related to the 2020 U.S. presidential elections?
>
> **Research Question 2**: Are there differences in the results provided by the same search engine to identical users under the default conditions?
>
> **Research Question 3**: Do the levels of discrepancies vary between searches about different political actors in relation to the 2020 U.S. presidential elections?

**Research Question 4**: Are there differences in the types of information sources prioritized by search engines for queries about different political candidates?

**Related Work: Algorithmic Impact Auditing and Political Search Results**

Algorithmic systems are essential elements of digital platform infrastructure. The need to assess their performance led to the formation of the set of methods collectively known as algorithmic auditing that is "a process of investigating the functionality and impact of decision-making algorithms" (Mittelstadt, 2016, P. 4994). While functionality auditing examines how algorithms arrive at certain decisions and outputs, impact auditing aims to find out which algorithmic outputs are prevalent and infer whether these outputs are biased in some way (Kroll et al., 2017; Sandvig et al., 2014). Algorithmic impact auditing of search engines is of paramount importance because they influence citizens' political information-seeking behavior by filtering and ranking politics-related information (Trevisan et al., 2018).

Since differences in search output can shift the opinions of undecided voters (Epstein & Robertson, 2015), biases in political search results can affect election outcomes and the general political landscape. In recent years, a number of studies that used algorithmic auditing in the context of political searches was conducted. Methodologically, such studies fall into three categories: those that rely on manually generated data (i.e., the ones collected from individual users or generated by the researchers themselves by manually querying search engines), those that rely on virtual agents simulating users' browsing behavior to generate and collect the data, and those that combine these two approaches.

The studies, which use manually generated data, primarily look on the effects of search personalization in the context of information behavior. Two of these studies investigated political filter bubbles on Google using crowd-sourced search results and found no evidence of bubbles' presence (Courtois et al., 2018; Robertson, Jiang, et al., 2018). Still, using a similar methodology another study found significant differences in personalized search results related to the U.S. presidential elections of 2016 (Robertson, Lazer, et al., 2018). Finally, a study that relied on the manual collection of the data by the researchers has assessed the diversity of search results in response to politically salient queries in the German context (Steiner et al., 2020). According to the findings, a certain degree of diversity is present even for the top results (depending on the query), but diversity generally increases for the long tail of search results.

The growing number of studies uses virtual agents to audit algorithmic content curation by search engines. One of the earliest studies (Feuz et al., 2011) on personalization of search results simulated browsing behavior of three different information-seeking personas on Google. The researchers found that results are affected by personalization and the effect increases overtime; the longer the virtual personas used the search engine, the more different were the results. Another study examined Google search results in the context of 2017 federal elections in Germany (Unkel & Haim, 2019). Specifically, the study simulated browsing activity of five information-oriented personas and showed the prevalence of general news websites and resources controlled by political parties in the results. Another study used a single virtual agent to query Google for a set of political queries that are germane to different ideological groups (Democrats vs. Republicans as the study was conducted in the U.S. context) and assess whether search engine results can be biased by the searcher's political orientation (Trielli & Diakopoulos, 2019). Another study that used a single virtual agent (Kulshrestha et al., 2019) investigated bias in Google's search results during the 2016 U.S. presidential election primaries. The authors found that Google's results tend to be biased in the direction of a specific candidate's political leaning (i.e., those related to "Donald Trump" exhibit a slight conservative bias and those related to "Hillary Clinton"—a slight liberal bias).

Besides studies relying exclusively on manual or agent-based data collection, there is some research combining the two approaches. The first study that combined virtual agent-based testing with crowd-sourced data for search engine auditing examined effects of different factors on search personalization (Hannak et al., 2013). The authors used virtual agents to generate a set of nonpersonalized results and compared them with the personalized results obtained from actual users. The study found that personalization significantly affects search results on both Bing and Google that were examined. In another study (Puschmann, 2019), the author asked the users to install a plug-in that queried Google for political searches at regular time intervals thus mimicking users' behavior and isolating potential bias related to the differences in the time when the searches were performed. The analysis revealed discrepancies in the ways different German parties were represented in Google Search and Google News in the run-up to the 2017 German federal elections.

The mentioned studies, with the exception of the one by Hannak and colleagues (2013) and the one by Steiner and colleagues (2020), have focused on one search engine—Google—and did not compare potential differences in algorithmic information curation between the search engines. This is understandable because Google currently dominates the global search market with around 90% of the market share (Statcounter, 2020) and is the engine that is the most commonly used by the majority of Western users. However, other search engines should not be overlooked because they are still used by millions of users across the globe and in some cases dominate regional search markets (i.e., Baidu is the leader on the Chinese market, and Yandex has around 50% of the market share in Russia; Statcounter, 2020). Furthermore, including other engines in the analysis allows testing whether some of them exhibit more biases than others and check whether the choice of a search engine itself affects the quality of information a user is exposed. Therefore, the first contribution to the existing scholarship that we aim to do is to compare politics-related results obtained through the six most popular search engines worldwide (Statcounter, 2020).

Apart from the lack of comparative research on search engine performance, aforementioned studies tend to look at the effects of personalization on search results and potential biases stemming from different variables (i.e., ideological bias of the searchers). None of them, however, has explored for the inherent randomization and volatility of search results. As search engines constantly and continuously update the results, the results inevitably change all the time. Hannak and colleagues (2013) have acknowledged the existence of this effect and attempted to control for it in their study by adding a control virtual agent. However, it is unclear if adding a single control agent is enough—that is, if noise affects all identical results equally. In addition, the scope of the differences in search results due to continuous search updates and inherent randomization has not been extensively examined to date. The only evidence on the level influence of these effects on search outputs comes from a commercial tool that tracks the volatility of search results for the same user throughout the day (SEMrush, n.d.) and from a study that found significant differences in the results for a singular "coronavirus" query when executed by several identical users at the exact same time under the same default filtering and ranking conditions (Makhortykh et al., 2020). With the present study, we aim to partially address this gap by examining the effects of the continuous search updates on the results through a systematic comparison of the results across several search queries and engines.

**Method**

*Data Collection*

Using automated agents to simulate browsing behavior of Internet users, we collected the HTML search results from the six most popular search engines according to Statcounter (2020): Google, Bing, Yahoo, Baidu, Yandex, and DuckDuckGo. Extending the methodology adopted by Haim et al. (2017), we built a cloud-based infrastructure to set up a controlled environment that allowed us to isolate external factors (e.g., time or location) and block the effects of search engine's in-built randomization (Makhortykh et al., 2020). Thus, our methodology addresses a potential limitation of earlier algorithmic auditing studies that did not account for the randomization effects (Kulshrestha et al., 2019; Puschmann, 2019; Steiner et al., 2020; Trielli & Diakopoulos, 2019; Unkel & Haim, 2019). Although one study looked at this effect for a singular query ("coronavirus") in the context of COVID-19 (Makhortykh et al., 2020), none, to date, examined the influence of randomization in the context of political search results, which is a gap we aim to address.

To implement the study, we used a cloud-based infrastructure made of 100 CentOS virtual machines deployed via Amazon Elastic Compute Cloud (EC2) and located in the Frankfurt EC2 region. We chose this particular region outside of the United States because (1) we considered that the usage of any region inside the United States might introduce biases in search results due to geolocation-based personalization as Republicans and Democrats are not evenly distributed across states; (2) we did not have the resources to afford more than one EC2 geographic region to counteract this potential effect (e.g., by selecting one pro-Republican and one pro-Democratic region), further, at the time of the analysis, EC2 had no clusters available in pro-Republican states; and (3) we selected Frankfurt because it serves as a base for many international companies and has a high share of English-speaking population.

All the machines were t3a.medium Amazon EC2 instances based on AMD EPYC 7,000 series processors. Each machine had two CPUs, four gigabyte (GB) RAM, and 20 GB hard drive. Because the machines were generated using the same Centos-based Amazon machine image with the same set of software installed (e.g., same Centos packages and browser versions), they had the same hardware and software specifications. Besides, all machines were located in the same range of Internet protocol (IPs) performed identical searches at the same time. Hence, the searches were conducted in a fully controlled environment that accounted for potential factors that could have led to the discrepancies in search results (e.g., due to personalization). The only difference between the machines related to their unique IP addresses—though they all belonged to the same IP range provided by EC2 and should not have affected the results due to, that is, location-based personalization. We do acknowledge, however, that this is a limitation of the present study, and future research should investigate the potential effects of the said discrepancy.

Each virtual machine hosted two browsers: Firefox and Chrome. In each browser ("agent"), we installed two extensions: a tracker and a bot. The tracker collected metadata (e.g., time stamps) and the full HTML of all pages that were visited within the browser that were sent to an external storage server. The bot emulated user browsing behavior by searching query terms from the predefined list (which included terms "us elections," "joe biden," "donald trump," and "bernie sanders") and navigating through the search results. The queries were selected based on the event that the search was centered on. We opted for generic actor names (e.g., instead of actor names accompanied by descriptions) to retrieve the least biased results about the actors. The focus on the three aforementioned actors is explained by the fact that ahead of the primary elections, Joe Biden and Bernie Sanders were the major contenders for the Democratic nomination, and Donald Trump was then-incumbent President running for the reelection. We added a generic "us elections" term to get a broader overview of search results at the time, which would not be biased toward one of the candidates. In the present study, we entered identical queries into search engines without accounting for potential differences in the ways search engine algorithms handle multiword queries.We opted for this to achieve maximum consistency between the searches which we deemed necessary as our study is focused on impact auditing. Future studies that focus on functionality auditing of web search algorithms might investigate, however, how multiword queries are handled by different algorithms.

Table 1. The Total Number of Agents That Completed the Task per Search Engine and Browser.

| Browser | Baidu | Bing | DDG | Google | Yahoo! | Yandex |
|---|---|---|---|---|---|---|
| Firefox | 15 | 16 | 17 | 17 | 15 | 16/6 (*) |
| Chrome | 16 | 15 | 17 | 16 | 16 | 16/13 (*) |

**Note**. The first row displays the name of the search engine (DDG is an abbreviation for DuckDuckGo), and the first column shows the name of the browser. (*) ¼ We obtained fewer results for Yandex for the "U.S. elections" query because it triggered the bot detection algorithm of Yandex which blocked some of the agents.

The navigation through the retrieved search results was organized in browser sessions, which consisted of three steps: (1) visiting the main landing page of a search engine, (2) inputting a query from the predefined list into the search text box, and "clicking" on the search button, and (3) navigating through the search results.

Each agent collected at least the top 50 results by visiting multiple result pages or by scrolling down the page (in case of infinite scrolling configuration of the search result page, such as in the case of DuckDuckGo). Immediately after each search session, the browsers were cleaned to prevent previous searches from affecting the following sessions. The bot removed both the data accessed by the browser (i.e., browsing history and cache) and the browser data that can be retrieved by the search engines' algorithms (i.e., local storage, session storage, and cookies). At the time of the data collection, none of the search engines was forcing the users to accept or reject their cookie policies. Hence, none of the agents accepted engine-specific policies.

Regardless of the search engine, each search session lasted less than 3 min. Each subsequent session started 7 min after the beginning of the previous one to guarantee at least a 4-min gap between sessions. Therefore, the agents were always synchronized at the beginning of all sessions to isolate the potential effect of time on search results.

The 200 agents were deployed on February 26, 2020, 1 day after the Democratic debate and almost a week before Super Tuesday, when 14 states hold democratic primary elections. The search engines (Baidu, Bing, DuckDuckGo, Google, Yandex, and Yahoo) selected for this study were equally distributed among the agents, so that 32 of 33 agents (15 of 16 from each browser group) were assigned to each search engine. During our collection, the expected amount of agents was slightly decreased because of the issues: (1) bot detection in Yandex via occasionally appearing captchas, and (2) a few browser crashes due to the limited volume of RAM available on the machines. The total number of agents providing data for each browser–engine combination is provided below (Table 1).

*Data Analysis*

After collecting the data, we used BeautifulSoup (Python; Richardson, 2020.) and rvest (R; Wickham & RStudio, 2019) packages to extract search results from the HTML for each query and filter out the URLs not related to the search results (e.g., ads). The latter decision is explained by our implicit interest in the default mechanisms for search filtering and ranking, not in the ads displayed by the engines. Then, for each query, we compared the URLs of the search results obtained by each possible pair of agents. We used two similarity metrics—Jaccard Index (JI) and Rank Biased Overlap (RBO).

JI measures the overlap between two sets of results and shows the size of the intersection between the sets over the union. JI has been used to measure similarities in search results by previous studies personalization of web search (Hannak et al., 2013; Kliman-Silver et al., 2015; Puschmann, 2019). The values of the JI vary from 0 to 1, with 1 indicating that the compared sets are identical, and 0 that they are completely different.

Although JI is valuable for assessing the similarity between two sets of results, it does not take into account their ranking. Yet, the latter feature is especially relevant for the present study due to the proven effect of search ranking on voting preferences (Epstein & Robertson, 2015). For this reason, we also used RBO metric that accounts for the order in which results are presented and is frequently utilized in the studies on search engines (Cardoso & Magalhães, 2011; Robertson, Jiang, et al., 2018; Robertson, Lazer, et al., 2018). Specifically, RBO takes into consideration three important characteristics of web search: incompleteness (there are too many search results so it is not possible to scrape all of them), indefiniteness (chosen result range is arbitrary), and top-weightedness (variation between the top results is more important than the one between the lower ones) of the results (Webber et al., 2010). The formula for RBO is as follows:

$$\text{RBO}(S, T, p) = (1 - p) \sum_{d=1}^{\infty} p^{d-1} \cdot A_d,$$

where *S* and *T* are two infinite rankings, d is the depth to which their agreement is computed, A is the level of agreement (which is equal to a Jaccard similarity of top d results), and the persistence parameter p determines the importance of top results: The lower the value of p, the more weight is assigned to the top results.

For each of the four queries, we calculated JI for the overall set of results and for top 10 results, and RBO (p = .95 and p = .8) for all the result pairs. Setting p to .95 allowed us to conduct a more systemic analysis of the differences between result pairs, whereas p = .8 enabled us to put more emphasis on the first few results (Webber et al., 2010).

After calculating JI and RBO, we aggregated the data for each search engine examined in the study. To make sure that the agents' browsers did not cause the discrepancies between the search results, we aggregated data separately for Chrome and for Firefox. This also allowed us to check whether search results differ between the two browsers for otherwise identical agents. Afterward, we calculated the mean values for JI and RBO between the sets of agents with different combinations of search engines and browsers they were produced by.

We produced a linear mixed effect model using the lme4 and lmerTest R packages (Bates et al., 2020; Kuznetsova et al., 2020) to fit the data and calculate the statistical significance for our main independent variables: browser and search engine. The model allows us to control for (1) the effects of multiple comparisons of each agent, that is, the search results of an agent are used several times, one per comparison against the search results of the other agents, and (2) the effects of the machine combination, that is, since each machine contains two agents (one per browser), we need to control for pairing the same machines several times.

To assess whether there are qualitative differences in the types of content prioritized by the search engines in response to different queries, we have first aggregated data about the domains that appeared most frequently in the top 20 results for each search engine–query combination. Then, we have manually coded the results based on the following categories:

- think tank/academic websites (i.e., academic articles/think tank reports),
- social media sites (i.e., Facebook, Twitter),
- reference work (i.e., dictionaries, encyclopedic notes, Wikipedia),
- news aggregators (i.e., *Google News*),
- legacy media (i.e., *New York Times*),
- infotainment (i.e., soft news websites such as Buzzfeed),
- government (i.e., White House website),
- fact-checking websites (i.e., PolitiFact),
- commerce (i.e., online shops),
- campaign (i.e., official candidate–affiliated campaign websites),
- alternative media (i.e., digital-born partisan outlets such as Conservapedia), and

- not available (i.e., the link points to a site/page that is no longer available at the time of the analysis).

The coding was performed by one of the authors and then thoroughly checked by the two other authors to ensure agreement between them. All the disagreements arising from the checks were resolved via consensus-coding in a series of group discussions.

To answer our research questions, we first looked at the differences in the results obtained for the "us elections" query via different search engines (e.g., Google vs. Yahoo; Research Question 1), then by different agents using the same browser (e.g., when both agents used Google; Research Question 2). Then, we repeated these two steps for the queries related to specific politicians (i.e., "bernie sanders," "donald trump," "joe biden") and checked the discrepancies between the results obtained for each query (Research Question 3). Finally, we have qualitatively examined and categorized the types of domains in top 20 results for different search engines and analyzed the differences in the types of sources prioritized by each engine for specific politicians (Research Question 4).

**Results**

*Differences in Search Results on "us elections" Query*

In response to Research Question 1, we find significant discrepancies between filtering and ranking mechanisms utilized by different search engines for the "us elections" query on both, Chrome and Firefox browsers (see Supplemental Material for the complete statistical summary). We observe that 115 of 120 (95.8%) similarity values between different search engines are lower than .35, including all the JI values for the top 10 search results (Figure 1). The ranking of the results is also highly volatile that means even in the nonpersonalized setting, users of the same search engine are unlikely to see the same results. While some discrepancies in the results provided by different search engines are expected, given that they utilize different algorithms to filter and rank results, the magnitude of discrepancies suggests that users of these platforms get fundamentally different sets of information.

The most similar results are provided by DuckDuckGo and Yahoo search engines. However, in this case, the two engines share just under half of all the results (n *50) results (measured by JI overall) and around a third of the top 10 results (JI for top 10). The similarities are even lower when the ranking is taken into account (RBO) with the ordering of top results (RBO, p = .8) in most cases being more volatile than the overall ordering of the results (RBO, p = .95). This finding echoes that of Steiner and colleagues (2020) who found that the differences in search results are higher for the lower positioned results. Still, for the second most similar pair, Bing and Yandex, the top results are more similar in terms of both, order and composition as indicated by higher top 10 JI and RBO with p = .8. Hence, the findings regarding the volatility of search results with different rankings are contextual.

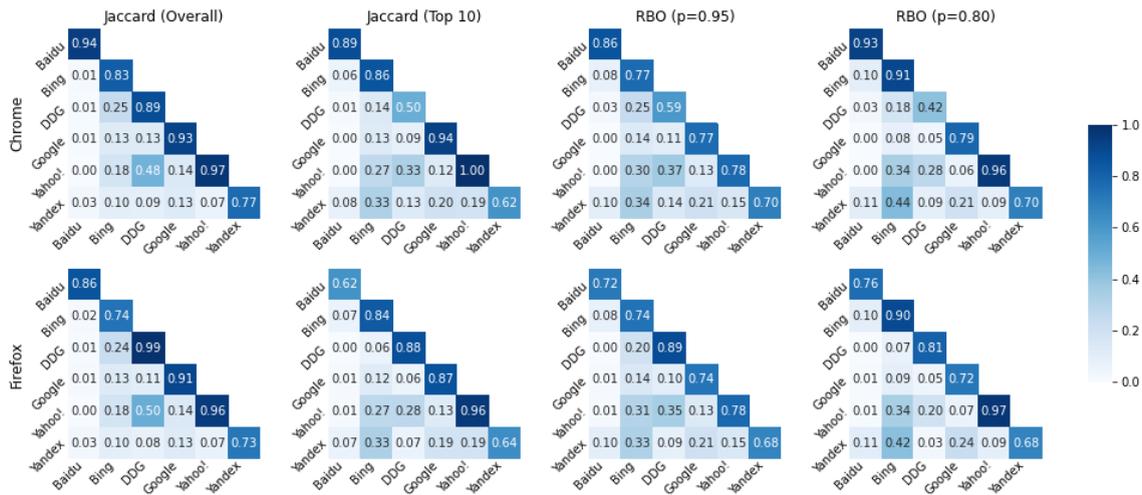

**Figure 1.** Cross-engine and cross-browser similarities in search results for the "us elections" search query. The columns show the different similarity measures used, and the rows show the results for Chrome (top) and Firefox (bottom) browsers.

The discrepancies in the results between different search engines can have important consequences for the public sphere because the users of different engines get different (political) information. However, most Western markets, including the United States, are currently dominated by Google. The Google's share of the U.S. search market is estimated to be just under 90%, slightly lower than Google's market share worldwide (Statcounter, 2020). This means that even if the results provided by Google are very different from those on other platforms, it does not affect about 90% of the U.S. public. However, what does affect the Western public is the high degree of randomization that creates discrepancies in the information curation even in the nonpersonalized context.

Concerning Research Question 2, we observed variations in search results within the same search engine for both Chrome and Firefox browsers (diagonal values in the plots in Figure 1). The only search engine in our sample did not randomize the selection of the top 10 results for the "us elections" query was Yahoo accessed from Chrome (but not from Firefox), and even in this case, the ordering of the results showed some variation between the agents. Since such variation happens under the nonpersonalized conditions, it is seemingly random and, most likely, attributed to the fact that search engine algorithms constantly adapt their output to provide the results viewed as the most relevant to the users at the given time. This constant output adaptation means that users of the same search engine are likely to receive different results even when they conduct searches at the same time and no personalization is involved. Even though the within-engine discrepancies are not as high as cross-engine ones, their effect on the public opinion is still important because the ranking of the search results can shift voters' opinion (Epstein & Robertson, 2015).

In terms of the browser differences, we only found significant differences between Firefox and Chrome for DuckDuckGo (the statistical table is reported in Supplemental Material, Table A1). In this case, the search results obtained in Firefox are more consistent than those obtained in Chrome for all our response variables. One potential explanation of this is that DuckDuckGo's search algorithm takes a user's browser into account when making curation decisions. If true, this is problematic since browser is a semantically nonmeaningful signal and its influence on search results can increase information inequalities between users of different browsers.

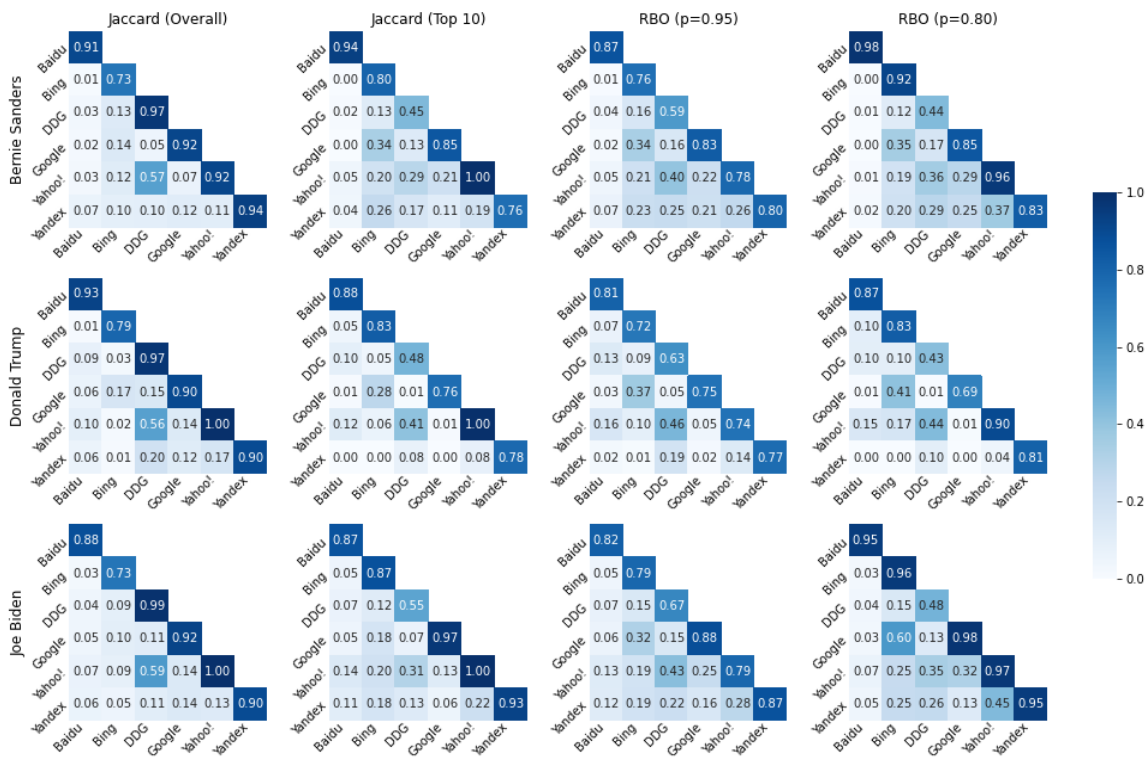

**Figure 2.** Similarities in search results across engines by relevant political candidate query, Chrome browser.

*Discrepancies in the Stability of Search Results for Different Political Candidates*

To answer Research Question 3, we compared the differences in the results for different political candidate queries. Similarly to the search results for the "us elections" query, there are large discrepancies for the queries related to specific political candidates (see Figure 2 for Chrome and Figure 3 for Firefox). On average, cross-engine dissimilarities for all the political candidates–related queries are slightly higher than those for the "us elections" query.

We found significant differences between the browsers depending on the query (statistically significant values are reported in Table 2; for the complete statistical tables see Supplemental Material, see Sections B.1, C.1, D.1). The results are in line with the findings for the "us elections" query: DuckDuckGo shuffles the results for Chrome users more than it does for Firefox users. For the "bernie sanders" query, we also observed some browser differences for Bing and Yahoo.

Looking only at the results for the same-engine comparisons and controlling for different browsers, we found statistically significant differences in terms of consistency of search results between the three queries (see Supplemental Material, Section E). These differences depend on the search engine that is being used, but overall "joe biden"- and "bernie

sanders"-related results are less volatile than "donald trump" ones in terms of JI (top 10), RBO (p ¼ .8) and RBO (p ¼ .95; see Supplemental Material, Section F).

*Prioritization of Source Types*

In order to infer qualitative differences between the results for queries on different political candidates and, thus, answer Research Question 4, we examined the top 20 results most frequently obtained through each engine for each of the three candidate-related queries (see Figures 4–6). We found that search engines, in general, prioritize different categories of search results, and in some cases (i.e., Baidu, Yahoo, and Yandex), there are large discrepancies between different candidate queries.

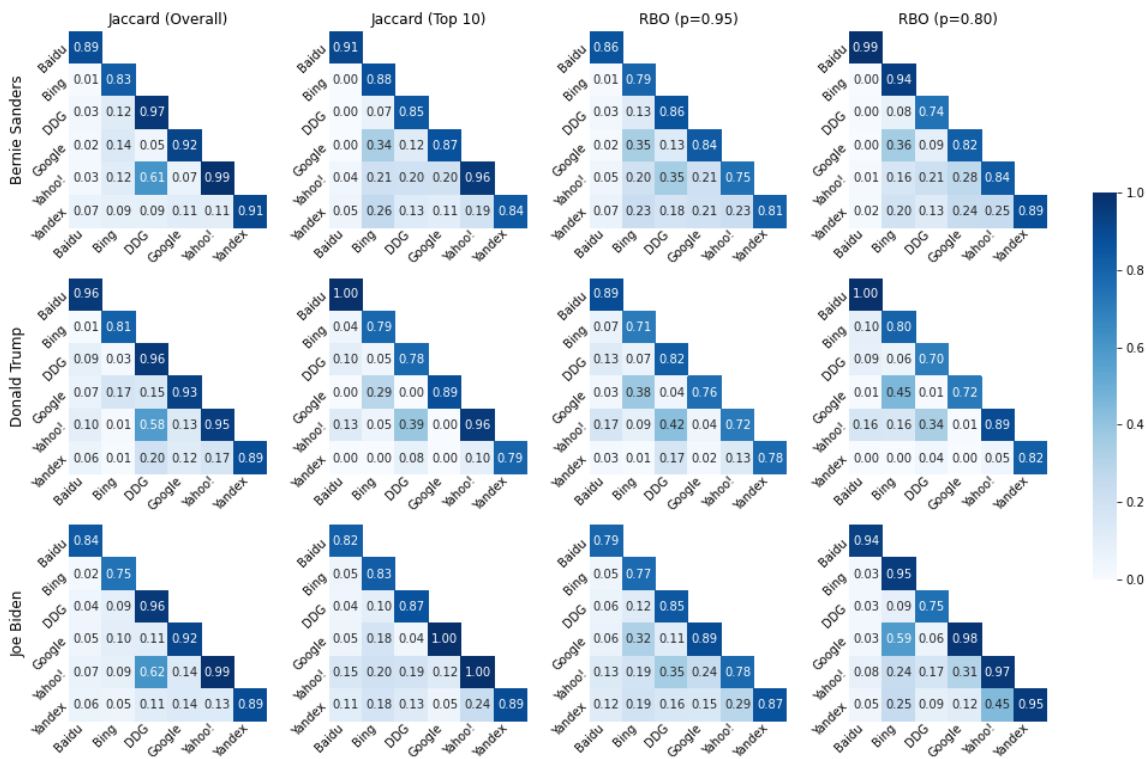

**Figure 3.** Similarities in search results across engines by relevant political candidate query, Firefox browser.

Table 2. p Values for Statistically Significant Effects of Browser in Politician Queries.

| Search query | Engine | JI (Overall) | JI (Top 10) | RBO (p = .8) | RBO ( p= .95) |
|---|---|---|---|---|---|
| bernie sanders | Bing | .016 | — | — | — |
| | DDG | — | <.0001 | .0296 | .0001 |
| | Yahoo | — | — | .0303 | — |
| joe biden | DDG | — | .0035 | .0002 | .0036 |

| donald trump | — | — | — | — | — |

**Note.** The first column shows the query term used. The second column refers to the search engine. JI = Jaccard Index; RBO = Rank Biased Overlap.

On Google, the prioritization of results is consistent for all three candidates with legacy media dominating search outputs. This is similar to what was observed in a study on Google search results in relation to the 2017 German federal elections (Unkel & Haim, 2019). Overall, legacy media results were more prevalent in the outputs of Google and Bing than those of the other search engines. The only major difference we observed on Google for different queries is that the results for "bernie sanders" did not contain a link to Sanders' campaign website, unlike those for "donald trump" and "joe biden." In addition, we found that the candidates-controlled campaign websites were less prevalent in Google results during the 2020 primaries than during the 2016 primaries when almost a quarter of top 10 results were made of candidate-affiliated websites (Kulshrestha et al., 2019). Still further longitudinal studies are necessary to properly verify this claim. As we conducted a snapshot experiment, we cannot state whether how persistent and systematic this observation is.

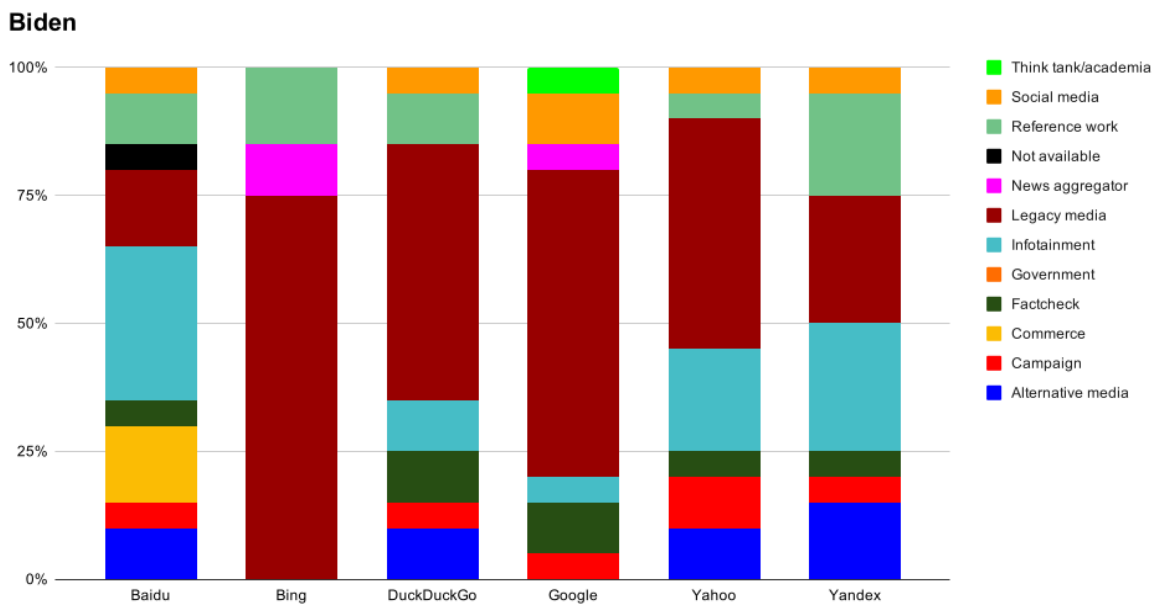

**Figure 4.** Information sources referenced in top 20 search results for "joe biden."

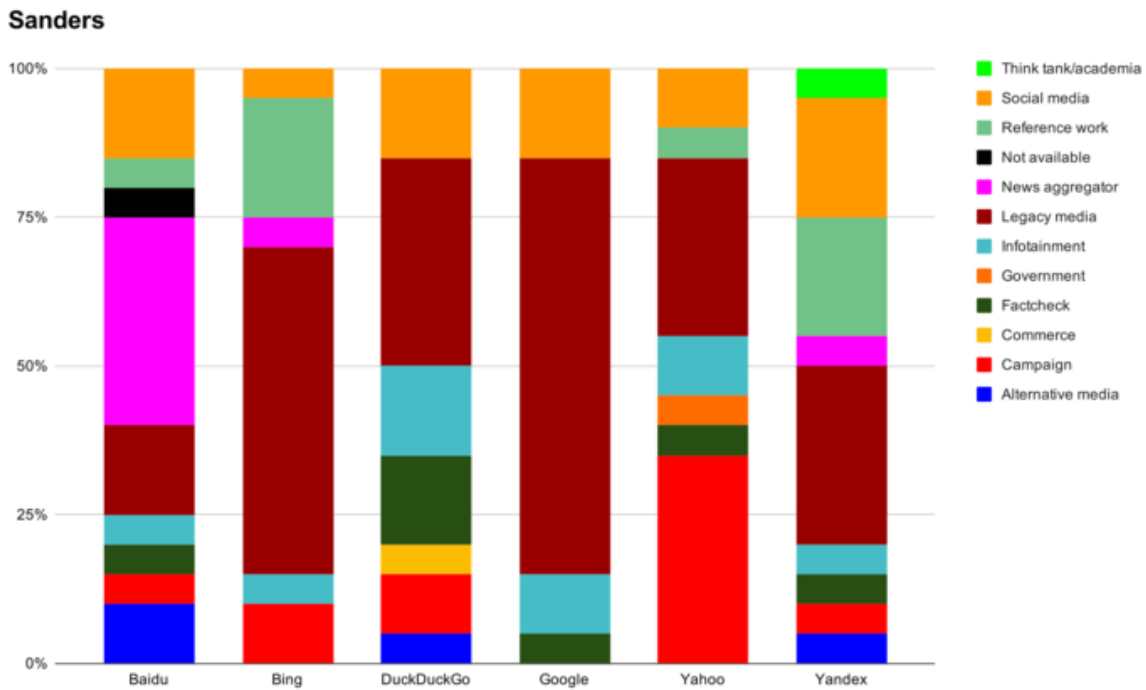

**Figure 5.** Information sources referenced in top 20 search results for "bernie sanders".

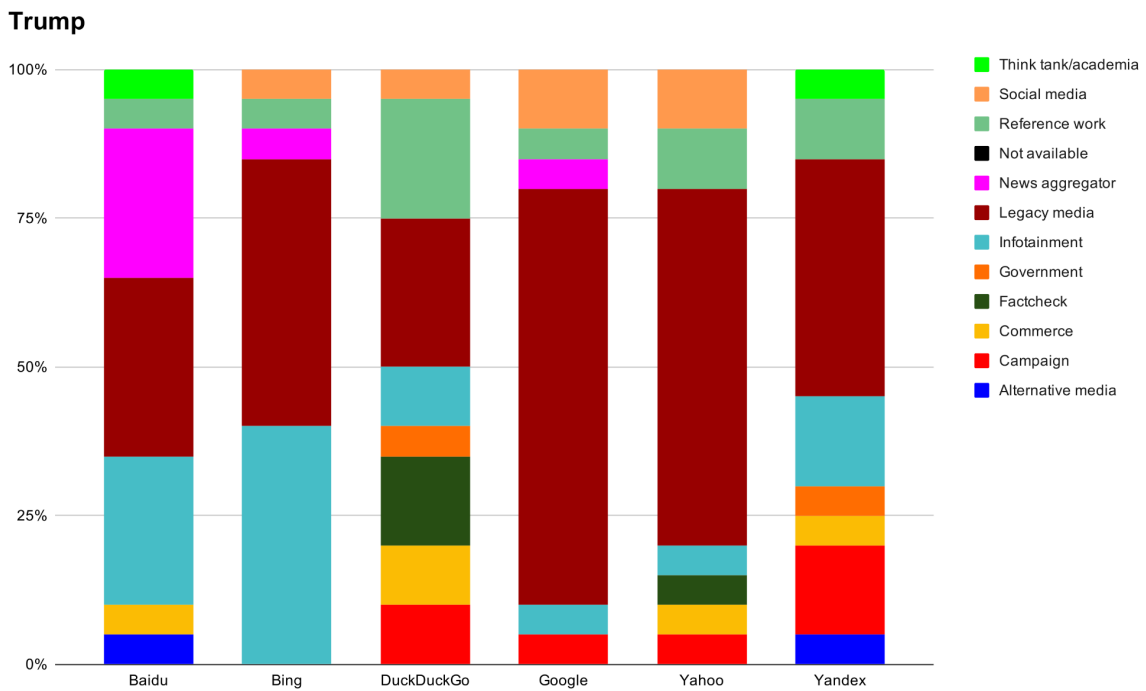

**Figure 6.** Information sources referenced in top 20 search results for "donald trump".

More pronounced differences across candidates were observed on Bing and Yahoo, the second and third most popular engines in the United States with 6.55% and 3.65% of the search market, respectively (Statcounter, 2020). In terms of potential biases, Yahoo displayed a high ratio of pro-Sanders results with around 40% of the top 20 results linking to outlets related to his campaign, while for Biden and Trump the ratio was 10% and 5%, respectively. However, as we conducted a snapshot experiment, we cannot say how stable the observed effect is overtime.

**Discussion**

Our findings highlight two major issues related to the ways search engine filter and rank political information. The first issue is a number of differences in the search outputs produced by algorithmic curation mechanisms of different search engines (Research Question 1). While some variation in the selection of information sources is expected as the engines clearly employ different algorithms to retrieve and rank the results, our study indicates that even under the nonpersonalized conditions, search results show varying degrees of volatility and prioritize different types of sources depending on the query. These discrepancies, even if they are to be expected due to the differences in algorithms, can lead to information inequalities between the individuals who use different search engines, in particular, as some of the engines seem to prioritize sources which are more supportive (e.g., Yahoo for Sanders) or critical (e.g., Yandex for Biden) of specific candidates. We suggest that this observation warrants further studies into how usage of different search engines affects the populations from the social science perspective.

While the effects of these inequalities might be somehow limited in the United States where over 90% of the public are using Google as their default engine (Statcounter, 2020), in the countries where the search market is not dominated by a single engine, the cross-engine discrepancies can have a larger effect on the public sphere. Such contexts include, for instance, East Asian states such as China, Japan, and South Korea, as well as post-Soviet countries such as Russia, Kazakhstan, and Belarus, where local search corporations serve as major competitors for Western tech giants (Statcounter, 2020).

The second troubling issue is the volatility of search results within the same search engine (RQ2). The randomization of search results is not necessarily a negative phenomenon, because it allows the Figure 6. Information sources referenced in top 20 search results for "donald trump." engines to present the most relevant information by updating the ranking of sources and can potentially diversify users' information diets (Helberger et al., 2018). On the other hand, such volatility makes search outputs less predictable and might lead to information inequalities between the users of the same engine by randomizing their access to information. We also find that the volatility of search results differs across different candidate queries (RQ3), with the results related to the two Democratic candidates being more stable than those for "donald trump" query.

In addition, our analysis has revealed qualitative differences in the composition of top results across the three political candidate queries (RQ4). For instance, we observed that Yahoo contained a much higher share of campaign websites for the "bernie sanders" query compared to other engines and queries. Such discrepancy might indicate a potential pro-Sanders bias in the output, but without a longitudinal study, it is not possible to verify how systematic this bias is. Further longitudinal research utilizing similar methodology is required to enhance our understanding of how resilient the observations coming from the current study are as we conducted a snapshot experiment and cannot state whether our observations indicate a presence of a systematic bias. Still, the observed differences in search results across political queries, engines and browsers are already troubling, because the ranking of political search results can affect voters' decisions (Epstein & Robertson, 2015).

In contrast to earlier research focusing on the effects of personalization on political information dissemination via search engines (i.e. Hannak et al., 2013; Puschmann, 2019; Unkel & Haim, 2019), our study highlights the need for taking into account search results' volatility that is present on all search engines we audited. Whereas personalization does not significantly alter election-related search results, at least on Google in the context of the German Federal elections (Unkel & Haim, 2019), our findings show that built-in randomization can strongly affect the composition and the ranking of results. It prompts the need to go beyond the current scholarship's focus on search personalization and its influence on promotion of specific biases (e.g., the ones related to gender and race; Noble, 2018) and discuss to what degree inherent volatility of the results can create informational inequalities which make users receive different information under identical conditions. Similar to the search queries related to the emergencies like the COVID-19 pandemic (Makhortykh et al., 2020), in the case of political queries such randomization, can result in some part of the population being less informed or even misinformed about important societal developments. Whether the users get to see certain information or not becomes, thus, a matter of chance that is in stark contradiction with the public's general perception of search results as accurate and trustworthy ("2020 Edelman Trust Barometer," n.d.; Pan et al., 2007) as well as the framing of the search process as unbiased and scientific by the search companies (Sweeney, 2013).

**Funding**
The authors disclosed receipt of the following financial support for the research, authorship, and/or publication of this article: This paper has been written within the project "Reciprocal relations between populist radica-lright attitudes and political information behaviour: A longitudinal study of attitude development in high-choice information environments" lead by Silke Adam (U of Bern) and Michaela Maier (U of Koblenz-Landau) and sponsored by the SNF (100001CL_182630/1) and DFG (MA 2244/9-1).

# Supplementary material

# Table of Contents

# A. Statistical Tests for "us elections" query

## A.1 Comparison of Browsers for "us elections" query

### Response: Jaccard

```
Linear mixed model fit by REML. t-tests use Satterthwaite's method ['lmerModLmerTest']
Formula: jaccard ~ browser * engine + (1 | agent1) + (1 | agent2) + (1 |      machine_combination)
   Data: subdf

REML criterion at convergence: -5752.8

Scaled residuals:
     Min      1Q  Median      3Q     Max
-13.1561 -0.2648  0.0226  0.1494  8.7211

Random effects:
 Groups              Name        Variance  Std.Dev.
 machine_combination (Intercept) 5.429e-05 0.007368
 agent2              (Intercept) 5.679e-03 0.075356
 agent1              (Intercept) 4.453e-03 0.066734
 Residual                        2.090e-03 0.045712
Number of obs: 2038, groups:  machine_combination, 1495; agent2, 179; agent1, 93

Fixed effects:
                            Estimate Std. Error         df t value Pr(>|t|)
(Intercept)                9.440e-01  2.556e-02  2.448e+02  36.932  < 2e-16 ***
browserFirefox            -4.868e-02  2.761e-02  1.679e+02  -1.763  0.07971 .
engineBing                -1.163e-01  3.676e-02  2.451e+02  -3.164  0.00175 **
engineDDG                 -5.418e-02  3.559e-02  2.442e+02  -1.522  0.12920
engineGoogle              -1.536e-02  3.615e-02  2.449e+02  -0.425  0.67133
engineYahoo!               2.676e-02  3.620e-02  2.460e+02   0.739  0.46052
engineYandex              -1.700e-01  3.827e-02  2.471e+02  -4.442 1.35e-05 ***
browserFirefox:engineBing  8.868e-04  3.907e-02  1.683e+02   0.023  0.98192
browserFirefox:engineDDG   9.801e-02  3.815e-02  1.675e+02   2.569  0.01106 *
browserFirefox:engineGoogle 4.323e-02 3.847e-02  1.681e+02   1.124  0.26272
browserFirefox:engineYahoo! 4.168e-02 3.910e-02  1.686e+02   1.066  0.28806
browserFirefox:engineYandex 3.888e-02 4.696e-02  1.686e+02   0.828  0.40877
---
Signif. codes:  0 '***' 0.001 '**' 0.01 '*' 0.05 '.' 0.1 ' ' 1

Correlation of Fixed Effects:
            (Intr) brwsrF engnBn engDDG engnGg engnY! engnYn brwF:B bF:DDG brwF:G brF:Y!
```

```
browserFrfx      -0.531
engineBing       -0.695  0.369
engineDDG        -0.718  0.382  0.499
engineGoogl      -0.707  0.376  0.492  0.508
engineYaho!      -0.706  0.375  0.491  0.507  0.499
engineYandx      -0.668  0.355  0.465  0.480  0.472  0.472
brwsrFrfx:B       0.375 -0.707 -0.540 -0.270 -0.265 -0.265 -0.251
brwsrFr:DDG       0.385 -0.724 -0.267 -0.535 -0.272 -0.272 -0.257  0.511
brwsrFrfx:G       0.381 -0.718 -0.265 -0.274 -0.540 -0.269 -0.255  0.507  0.520
brwsrFrf:Y!       0.375 -0.706 -0.261 -0.269 -0.265 -0.533 -0.251  0.499  0.511  0.507
brwsrFrfx:Y       0.312 -0.588 -0.217 -0.224 -0.221 -0.221 -0.469  0.415  0.426  0.422  0.415
convergence code: 0
Model failed to converge with max|grad| = 0.00216689 (tol = 0.002, component 1)
```

## Response: Jaccard Top-10

```
Linear mixed model fit by REML. t-tests use Satterthwaite's method ['lmerModLmerTest']
Formula: jacctop10 ~ browser * engine + (1 | agent1) + (1 | agent2) +      (1 | machine_combination)
   Data: subdf

REML criterion at convergence: -2107.7

Scaled residuals:
    Min      1Q  Median      3Q     Max
-4.8403 -0.3722  0.0307  0.2115  9.8368

Random effects:
 Groups              Name        Variance  Std.Dev.
 machine_combination (Intercept) 0.0009883 0.03144
 agent2              (Intercept) 0.0239715 0.15483
 agent1              (Intercept) 0.0224131 0.14971
 Residual                        0.0125251 0.11192
Number of obs: 2038, groups:  machine_combination, 1495; agent2, 179; agent1, 93

Fixed effects:
                            Estimate Std. Error        df t value Pr(>|t|)
(Intercept)                  0.89139    0.05500 235.84248  16.206  < 2e-16 ***
browserFirefox              -0.15870    0.05726 168.74932  -2.771  0.00621 **
engineBing                  -0.03403    0.07911 236.28394  -0.430  0.66749
engineDDG                   -0.38616    0.07657 235.05204  -5.044 9.14e-07 ***
engineGoogle                 0.04351    0.07779 235.93328   0.559  0.57652
engineYahoo!                 0.10865    0.07793 237.10221   1.394  0.16457
engineYandex                -0.27136    0.08241 238.76236  -3.293  0.00114 **
browserFirefox:engineBing    0.15514    0.08102 169.09470   1.915  0.05719 .
browserFirefox:engineDDG     0.29870    0.07908 168.18859   3.777  0.00022 ***
browserFirefox:engineGoogle  0.12862    0.07978 169.04424   1.612  0.10878
browserFirefox:engineYahoo!  0.13656    0.08112 169.46715   1.683  0.09413 .
browserFirefox:engineYandex  0.17570    0.09741 169.72004   1.804  0.07305 .
---
Signif. codes:  0 '***' 0.001 '**' 0.01 '*' 0.05 '.' 0.1 ' ' 1

Correlation of Fixed Effects:
            (Intr) brwsrF engnBn engDDG engnGg engnY! engnYn brwF:B bF:DDG brwF:G brF:Y!
browserFrfx -0.516
engineBing  -0.695  0.359
engineDDG   -0.718  0.371  0.499
engineGoogl -0.707  0.365  0.492  0.508
```

```
engineYaho! -0.706  0.364  0.491  0.507  0.499
engineYandx -0.667  0.344  0.464  0.479  0.472  0.471
brwsrFrfx:B  0.365 -0.706 -0.525 -0.262 -0.258 -0.257 -0.243
brwsrFr:DDG  0.373 -0.724 -0.260 -0.519 -0.264 -0.264 -0.249  0.511
brwsrFrfx:G  0.370 -0.718 -0.257 -0.266 -0.524 -0.261 -0.247  0.507  0.520
brwsrFrf:Y!  0.364 -0.706 -0.253 -0.262 -0.257 -0.518 -0.243  0.498  0.511  0.507
brwsrFrfx:Y  0.303 -0.588 -0.211 -0.218 -0.214 -0.214 -0.457  0.415  0.426  0.422  0.415
```

## Response: RBO (p=0.95)

```
Linear mixed model fit by REML. t-tests use Satterthwaite's method ['lmerModLmerTest']
Formula: rbo_95 ~ browser * engine + (1 | agent1) + (1 | agent2) + (1 |      machine_combination)
   Data: subdf

REML criterion at convergence: -3948.7

Scaled residuals:
    Min      1Q  Median      3Q     Max
-4.9900 -0.3979  0.0240  0.2390  8.3798

Random effects:
 Groups              Name        Variance  Std.Dev.
 machine_combination (Intercept) 0.0005483 0.02342
 agent2              (Intercept) 0.0076564 0.08750
 agent1              (Intercept) 0.0095810 0.09788
 Residual                        0.0050011 0.07072
Number of obs: 2038, groups:  machine_combination, 1495; agent2, 179; agent1, 93

Fixed effects:
                              Estimate Std. Error        df t value Pr(>|t|)
(Intercept)                    0.86423    0.03360 214.70327  25.720  < 2e-16 ***
browserFirefox                -0.08288    0.03262 167.93736  -2.541  0.01195 *
engineBing                    -0.09407    0.04833 215.15770  -1.947  0.05289 .
engineDDG                     -0.26817    0.04677 213.92278  -5.734 3.32e-08 ***
engineGoogle                  -0.09685    0.04752 214.79434  -2.038  0.04278 *
engineYahoo!                  -0.08466    0.04761 215.87886  -1.778  0.07677 .
engineYandex                  -0.16680    0.05036 217.62644  -3.312  0.00108 **
browserFirefox:engineBing      0.07050    0.04614 168.19459   1.528  0.12840
browserFirefox:engineDDG       0.19350    0.04504 167.24969   4.297 2.93e-05 ***
browserFirefox:engineGoogle    0.06788    0.04545 168.30664   1.494  0.13714
browserFirefox:engineYahoo!    0.08497    0.04622 168.67352   1.838  0.06776 .
browserFirefox:engineYandex    0.08757    0.05551 169.16331   1.578  0.11650
---
Signif. codes:  0 '***' 0.001 '**' 0.01 '*' 0.05 '.' 0.1 ' ' 1

Correlation of Fixed Effects:
            (Intr) brwsrF engnBn engDDG engnGg engnY! engnYn brwF:B bF:DDG brwF:G brF:Y!
browserFrfx -0.484
engineBing  -0.695  0.336
engineDDG   -0.718  0.348  0.500
engineGoogl -0.707  0.342  0.492  0.508
engineYaho! -0.706  0.341  0.491  0.507  0.499
engineYandx -0.667  0.323  0.464  0.479  0.472  0.471
brwsrFrfx:B  0.342 -0.705 -0.492 -0.246 -0.242 -0.241 -0.228
brwsrFr:DDG  0.350 -0.724 -0.244 -0.487 -0.248 -0.247 -0.234  0.511
brwsrFrfx:G  0.347 -0.718 -0.241 -0.249 -0.491 -0.245 -0.232  0.506  0.520
brwsrFrf:Y!  0.341 -0.706 -0.237 -0.245 -0.241 -0.486 -0.228  0.498  0.511  0.507
brwsrFrfx:Y  0.284 -0.588 -0.198 -0.204 -0.201 -0.201 -0.429  0.415  0.426  0.422  0.415
```

## Response:  RBO (p=0.8)

```
Linear mixed model fit by REML. t-tests use Satterthwaite's method ['lmerModLmerTest']
Formula: rbo_80 ~ browser * engine + (1 | agent1) + (1 | agent2) + (1 |      machine_combination)
   Data: subdf

REML criterion at convergence: -1574.3

Scaled residuals:
    Min     1Q  Median     3Q    Max
-4.4305 -0.4232  0.0458  0.2217  5.7322

Random effects:
 Groups              Name        Variance Std.Dev.
 machine_combination (Intercept) 0.002188 0.04678
 agent2              (Intercept) 0.011521 0.10734
 agent1              (Intercept) 0.016846 0.12979
 Residual                        0.017457 0.13212
Number of obs: 2038, groups:  machine_combination, 1495; agent2, 179; agent1, 93

Fixed effects:
                              Estimate Std. Error        df t value Pr(>|t|)
(Intercept)                    0.93227    0.04420 206.19512  21.094  < 2e-16 ***
browserFirefox                -0.09593    0.04184 167.94975  -2.293 0.023093 *
engineBing                    -0.01853    0.06361 207.24252  -0.291 0.771113
engineDDG                     -0.51262    0.06147 204.82107  -8.340 1.08e-14 ***
engineGoogle                  -0.14334    0.06252 206.36713  -2.293 0.022862 *
engineYahoo!                   0.02573    0.06266 207.47717   0.411 0.681720
engineYandex                  -0.23605    0.06647 211.76684  -3.551 0.000472 ***
browserFirefox:engineBing      0.09161    0.05920 168.36862   1.547 0.123650
browserFirefox:engineDDG       0.22999    0.05771 166.63698   3.986 0.000100 ***
browserFirefox:engineGoogle    0.06326    0.05833 168.71031   1.085 0.279687
browserFirefox:engineYahoo!    0.10260    0.05933 168.55464   1.729 0.085584 .
browserFirefox:engineYandex    0.12156    0.07135 170.67013   1.704 0.090231 .
---
Signif. codes:  0 '***' 0.001 '**' 0.01 '*' 0.05 '.' 0.1 ' ' 1

Correlation of Fixed Effects:
            (Intr) brwsrF engnBn engDDG engnGg engnY! engnYn brwF:B bF:DDG brwF:G brF:Y!
browserFrfx -0.487
engineBing  -0.695  0.338
engineDDG   -0.719  0.350  0.500
engineGoogl -0.707  0.344  0.491  0.508
engineYaho! -0.705  0.343  0.490  0.507  0.499
engineYandx -0.665  0.324  0.462  0.478  0.470  0.469
brwsrFrfx:B  0.344 -0.703 -0.497 -0.247 -0.243 -0.243 -0.229
brwsrFr:DDG  0.353 -0.725 -0.245 -0.489 -0.250 -0.249 -0.235  0.510
brwsrFrfx:G  0.349 -0.717 -0.243 -0.251 -0.494 -0.246 -0.232  0.504  0.520
brwsrFrf:Y!  0.343 -0.705 -0.239 -0.247 -0.243 -0.490 -0.228  0.496  0.511  0.508
brwsrFrfx:Y  0.286 -0.586 -0.198 -0.205 -0.202 -0.201 -0.436  0.412  0.427  0.421  0.414
```

## A.2 Comparison of Search Engines for "us elections" query

### Response:  Jaccard

Linear mixed model fit by REML. t-tests use Satterthwaite's method ['lmerModLmerTest']

```
Formula: jaccard ~ engines_ + (1 | agent1) + (1 | agent2) + (1 | browser) +      (1 | machine_combination)
   Data: subdf

REML criterion at convergence: -36722.3

Scaled residuals:
    Min      1Q  Median      3Q     Max
-9.8246 -0.3578 -0.0078  0.3606  5.3976

Random effects:
 Groups              Name        Variance  Std.Dev.
 machine_combination (Intercept) 0.000e+00 0.0000000
 agent2              (Intercept) 3.377e-05 0.0058115
 agent1              (Intercept) 3.388e-05 0.0058202
 browser             (Intercept) 1.596e-07 0.0003995
 Residual                        2.033e-04 0.0142590
Number of obs: 6637, groups:  machine_combination, 4005; agent2, 179; agent1, 179; browser, 2

Fixed effects:
                       Estimate Std. Error        df t value Pr(>|t|)
(Intercept)            0.013855   0.001267  10.399383  10.932 4.97e-07 ***
engines_Baidu-DDG     -0.003405   0.001368 800.323431  -2.489    0.013 *
engines_Baidu-Google  -0.001605   0.001378 800.620999  -1.165    0.244
engines_Baidu-Yahoo!  -0.011900   0.001400 801.679909  -8.501  < 2e-16 ***
engines_Baidu-Yandex   0.020623   0.001608 792.928943  12.821  < 2e-16 ***
engines_Bing-DDG       0.229403   0.001369 802.880103 167.514  < 2e-16 ***
engines_Bing-Google    0.114711   0.001377 799.011637  83.317  < 2e-16 ***
engines_Bing-Yahoo!    0.167255   0.001400 801.875053 119.462  < 2e-16 ***
engines_Bing-Yandex    0.084766   0.001613 809.568100  52.561  < 2e-16 ***
engines_DDG-Google     0.105032   0.001705 483.855589  61.594  < 2e-16 ***
engines_DDG-Yahoo!     0.479521   0.001722 487.987434 278.400  < 2e-16 ***
engines_DDG-Yandex     0.071928   0.001889 522.909088  38.069  < 2e-16 ***
engines_Google-Yahoo!  0.129410   0.001731 490.605103  74.758  < 2e-16 ***
engines_Google-Yandex  0.113973   0.001901 529.376300  59.967  < 2e-16 ***
engines_Yahoo!-Yandex  0.056281   0.001917 532.680918  29.353  < 2e-16 ***
---
Signif. codes:  0 '***' 0.001 '**' 0.01 '*' 0.05 '.' 0.1 ' ' 1
convergence code: 0
boundary (singular) fit: see ?isSingular
```

## Response: Jaccard Top-10

```
Linear mixed model fit by REML. t-tests use Satterthwaite's method ['lmerModLmerTest']
Formula: jacctop10 ~ engines_ + (1 | agent1) + (1 | agent2) + (1 | browser) +      (1 | machine_combination)
   Data: subdf

REML criterion at convergence: -26194.9

Scaled residuals:
    Min      1Q  Median      3Q     Max
-5.1249 -0.3260 -0.0321  0.2853  9.1889

Random effects:
 Groups              Name        Variance  Std.Dev.
 machine_combination (Intercept) 0.000e+00 0.000000
 agent2              (Intercept) 9.967e-04 0.031571
 agent1              (Intercept) 8.981e-04 0.029968
```

```
 browser           (Intercept) 7.884e-05 0.008879
 Residual                      9.108e-04 0.030179
Number of obs: 6637, groups:  machine_combination, 4005; agent2, 179; agent1, 179; browser, 2

Fixed effects:
                      Estimate Std. Error        df t value Pr(>|t|)
(Intercept)           0.061965   0.008478   2.044701   7.309   0.0171 *
engines_Baidu-DDG    -0.055757   0.005732 419.003649  -9.727  < 2e-16 ***
engines_Baidu-Google -0.052943   0.005773 418.890355  -9.171  < 2e-16 ***
engines_Baidu-Yahoo! -0.053319   0.005864 419.378036  -9.092  < 2e-16 ***
engines_Baidu-Yandex  0.013480   0.006755 420.515588   1.996   0.0466 *
engines_Bing-DDG      0.037483   0.005734 419.788063   6.537 1.83e-10 ***
engines_Bing-Google   0.062807   0.005773 419.013786  10.879  < 2e-16 ***
engines_Bing-Yahoo!   0.208913   0.005864 419.534429  35.625  < 2e-16 ***
engines_Bing-Yandex   0.264007   0.006752 421.835188  39.100  < 2e-16 ***
engines_DDG-Google    0.011498   0.007894 371.765103   1.456   0.1461
engines_DDG-Yahoo!    0.242587   0.007960 372.467799  30.475  < 2e-16 ***
engines_DDG-Yandex    0.039040   0.008628 378.626392   4.525 8.10e-06 ***
engines_Google-Yahoo! 0.064182   0.007991 372.939894   8.032 1.27e-14 ***
engines_Google-Yandex 0.133363   0.008657 379.534443  15.406  < 2e-16 ***
engines_Yahoo!-Yandex 0.124684   0.008723 380.285031  14.293  < 2e-16 ***
---
Signif. codes:  0 '***' 0.001 '**' 0.01 '*' 0.05 '.' 0.1 ' ' 1
convergence code: 0
boundary (singular) fit: see ?isSingular
```

## Response:  RBO (p=0.95)

```
Linear mixed model fit by REML. t-tests use Satterthwaite's method ['lmerModLmerTest']
Formula: rbo_95 ~ engines_ + (1 | agent1) + (1 | agent2) + (1 | browser) +     (1 | machine_combination)
   Data: subdf

REML criterion at convergence: -27931.4

Scaled residuals:
    Min      1Q  Median      3Q     Max
-4.3397 -0.4261 -0.0161  0.3647  5.3031

Random effects:
 Groups              Name        Variance  Std.Dev.
 machine_combination (Intercept) 1.403e-13 3.745e-07
 agent2              (Intercept) 6.204e-04 2.491e-02
 agent1              (Intercept) 5.408e-04 2.325e-02
 browser             (Intercept) 3.045e-05 5.518e-03
 Residual                        7.094e-04 2.664e-02
Number of obs: 6637, groups:  machine_combination, 4005; agent2, 179; agent1, 179; browser, 2

Fixed effects:
                      Estimate Std. Error        df t value Pr(>|t|)
(Intercept)           0.079702   0.005952   2.614099  13.392 0.001762 **
engines_Baidu-DDG    -0.066225   0.004552 438.151952 -14.550  < 2e-16 ***
engines_Baidu-Google -0.071770   0.004584 437.920822 -15.656  < 2e-16 ***
engines_Baidu-Yahoo! -0.071976   0.004657 438.678196 -15.456  < 2e-16 ***
engines_Baidu-Yandex  0.021282   0.005362 439.902935   3.969 8.42e-05 ***
engines_Bing-DDG      0.143178   0.004554 439.237961  31.440  < 2e-16 ***
engines_Bing-Google   0.063067   0.004584 438.293658  13.757  < 2e-16 ***
engines_Bing-Yahoo!   0.223475   0.004657 438.904286  47.987  < 2e-16 ***
```

```
engines_Bing-Yandex      0.254050   0.005361 441.838793  47.385  < 2e-16 ***
engines_DDG-Google       0.023378   0.006223 377.679596   3.757 0.000199 ***
engines_DDG-Yahoo!       0.279278   0.006275 378.571252  44.503  < 2e-16 ***
engines_DDG-Yandex       0.039026   0.006807 386.422319   5.734 1.98e-08 ***
engines_Google-Yahoo!    0.052152   0.006300 379.167096   8.278 2.16e-15 ***
engines_Google-Yandex    0.130837   0.006831 387.608162  19.154  < 2e-16 ***
engines_Yahoo!-Yandex    0.073975   0.006883 388.508537  10.747  < 2e-16 ***
---
Signif. codes:  0 '***' 0.001 '**' 0.01 '*' 0.05 '.' 0.1 ' ' 1
convergence code: 0
boundary (singular) fit: see ?isSingular
```

## Response:  RBO (p=0.8)

```
Linear mixed model fit by REML. t-tests use Satterthwaite's method ['lmerModLmerTest']
Formula: rbo_80 ~ engines_ + (1 | agent1) + (1 | agent2) + (1 | browser) +      (1 | machine_combination)
   Data: subdf

REML criterion at convergence: -18314.5

Scaled residuals:
    Min      1Q  Median      3Q     Max
-3.9570 -0.4259  0.0071  0.3608  4.2815

Random effects:
 Groups              Name        Variance  Std.Dev.
 machine_combination (Intercept) 0.0000000 0.00000
 agent2              (Intercept) 0.0014840 0.03852
 agent1              (Intercept) 0.0012271 0.03503
 browser             (Intercept) 0.0001537 0.01240
 Residual                        0.0031302 0.05595
Number of obs: 6637, groups:  machine_combination, 4005; agent2, 179; agent1, 179; browser, 2

Fixed effects:
                       Estimate Std. Error         df t value Pr(>|t|)
(Intercept)            0.101393   0.011273   1.882299   8.994 0.014620 *
engines_Baidu-DDG     -0.084542   0.007363 518.469900 -11.482  < 2e-16 ***
engines_Baidu-Google  -0.097356   0.007415 517.749237 -13.130  < 2e-16 ***
engines_Baidu-Yahoo!  -0.095693   0.007534 519.490938 -12.701  < 2e-16 ***
engines_Baidu-Yandex   0.006272   0.008680 520.270269   0.723 0.470216
engines_Bing-DDG       0.024038   0.007370 520.819008   3.262 0.001179 **
engines_Bing-Google   -0.011789   0.007416 519.094469  -1.590 0.112500
engines_Bing-Yahoo!    0.237549   0.007535 520.216997  31.525  < 2e-16 ***
engines_Bing-Yandex    0.331605   0.008684 524.188412  38.185  < 2e-16 ***
engines_DDG-Google    -0.053472   0.009795 401.633188  -5.459 8.39e-08 ***
engines_DDG-Yahoo!     0.136270   0.009882 403.296144  13.789  < 2e-16 ***
engines_DDG-Yandex    -0.037864   0.010760 416.956260  -3.519 0.000481 ***
engines_Google-Yahoo! -0.034742   0.009924 404.406091  -3.501 0.000516 ***
engines_Google-Yandex  0.116982   0.010804 419.356467  10.827  < 2e-16 ***
engines_Yahoo!-Yandex -0.011246   0.010893 420.778695  -1.032 0.302461
---
Signif. codes:  0 '***' 0.001 '**' 0.01 '*' 0.05 '.' 0.1 ' ' 1
convergence code: 0
boundary (singular) fit: see ?isSingular
```

# B. Statistical Tests for "bernie sanders" query

## B.1 Comparison of Browsers for "us elections" query

### Response: Jaccard

```
Linear mixed model fit by REML. t-tests use Satterthwaite's method ['lmerModLmerTest']
Formula: jaccard ~ browser * engine + (1 | agent1) + (1 | agent2) + (1 |      machine_combination)
   Data: subdf

REML criterion at convergence: -6102.4

Scaled residuals:
    Min      1Q  Median      3Q     Max
-3.7102 -0.4353  0.0197  0.2927 12.8798

Random effects:
 Groups              Name        Variance  Std.Dev.
 machine_combination (Intercept) 1.409e-05 0.003753
 agent2              (Intercept) 2.201e-03 0.046913
 agent1              (Intercept) 4.136e-03 0.064314
 Residual                        2.701e-03 0.051972
Number of obs: 2258, groups:  machine_combination, 1537; agent2, 192; agent1, 96

Fixed effects:
                            Estimate Std. Error         df t value Pr(>|t|)
(Intercept)                 0.909725   0.020517 189.589718  44.340  < 2e-16 ***
browserFirefox             -0.011024   0.017902 166.464863  -0.616   0.5389
engineBing                 -0.174685   0.029524 190.330285  -5.917  1.5e-08 ***
engineDDG                   0.060930   0.028548 188.652320   2.134   0.0341 *
engineGoogle                0.006559   0.029000 189.236797   0.226   0.8213
engineYahoo!                0.007027   0.029014 189.558048   0.242   0.8089
engineYandex                0.027131   0.029044 190.230746   0.934   0.3514
browserFirefox:engineBing   0.061620   0.025381 168.134029   2.428   0.0162 *
browserFirefox:engineDDG    0.010162   0.024699 165.353888   0.411   0.6813
browserFirefox:engineGoogle 0.013795   0.024929 166.571360   0.553   0.5807
browserFirefox:engineYahoo! 0.044427   0.025315 166.451331   1.755   0.0811 .
browserFirefox:engineYandex -0.006441  0.025154 167.298795  -0.256   0.7982
---
Signif. codes:  0 '***' 0.001 '**' 0.01 '*' 0.05 '.' 0.1 ' ' 1

Correlation of Fixed Effects:
            (Intr) brwsrF engnBn engDDG engnGg engnY! engnYn brwF:B bF:DDG brwF:G brF:Y!
browserFrfx -0.442
engineBing  -0.695  0.307
engineDDG   -0.719  0.318  0.499
engineGoogl -0.707  0.313  0.492  0.508
engineYaho! -0.707  0.313  0.491  0.508  0.500
engineYandx -0.706  0.312  0.491  0.508  0.500  0.500
brwsrFrfx:B  0.312 -0.705 -0.450 -0.224 -0.221 -0.221 -0.220
brwsrFr:DDG  0.321 -0.725 -0.223 -0.444 -0.227 -0.227 -0.226  0.511
brwsrFrfx:G  0.318 -0.718 -0.221 -0.228 -0.448 -0.225 -0.224  0.506  0.520
brwsrFrf:Y!  0.313 -0.707 -0.217 -0.225 -0.221 -0.442 -0.221  0.499  0.513  0.508
brwsrFrfx:Y  0.315 -0.712 -0.219 -0.226 -0.223 -0.223 -0.447  0.502  0.516  0.511  0.503
```

## Response: Jaccard Top-10

```
Linear mixed model fit by REML. t-tests use Satterthwaite's method ['lmerModLmerTest']
Formula: jacctop10 ~ browser * engine + (1 | agent1) + (1 | agent2) +      (1 | machine_combination)
   Data: subdf

REML criterion at convergence: -3045.4

Scaled residuals:
    Min      1Q  Median      3Q     Max
-3.8942 -0.5392  0.0001  0.4503  7.7916

Random effects:
 Groups              Name        Variance  Std.Dev.
 machine_combination (Intercept) 0.0006679 0.02584
 agent2              (Intercept) 0.0050696 0.07120
 agent1              (Intercept) 0.0188896 0.13744
 Residual                        0.0102972 0.10148
Number of obs: 2258, groups:  machine_combination, 1537; agent2, 192; agent1, 96

Fixed effects:
                             Estimate Std. Error        df t value Pr(>|t|)
(Intercept)                  0.941024   0.039962 148.325020  23.548  < 2e-16 ***
browserFirefox              -0.013000   0.028285 168.263530  -0.460  0.64639
engineBing                  -0.140327   0.057508 148.935605  -2.440  0.01586 *
engineDDG                   -0.491117   0.055605 147.587207  -8.832 2.77e-15 ***
engineGoogle                -0.090829   0.056488 148.068773  -1.608  0.10998
engineYahoo!                 0.058994   0.056512 148.307272   1.044  0.29823
engineYandex                -0.182246   0.056564 148.789937  -3.222  0.00156 **
browserFirefox:engineBing    0.052888   0.040104 169.854801   1.319  0.18902
browserFirefox:engineDDG     0.159345   0.038987 166.634955   4.087 6.78e-05 ***
browserFirefox:engineGoogle  0.019089   0.039393 168.596770   0.485  0.62861
browserFirefox:engineYahoo! -0.009252   0.039997 168.275026  -0.231  0.81736
browserFirefox:engineYandex  0.051879   0.039770 169.305894   1.304  0.19384
---
Signif. codes:  0 '***' 0.001 '**' 0.01 '*' 0.05 '.' 0.1 ' ' 1

Correlation of Fixed Effects:
            (Intr) brwsrF engnBn engDDG engnGg engnY! engnYn brwF:B bF:DDG brwF:G brF:Y!
browserFrfx -0.368
engineBing  -0.695  0.256
engineDDG   -0.719  0.265  0.499
engineGoogl -0.707  0.261  0.492  0.508
engineYaho! -0.707  0.260  0.491  0.508  0.500
engineYandx -0.706  0.260  0.491  0.508  0.500  0.500
brwsrFrfx:B  0.260 -0.703 -0.376 -0.187 -0.184 -0.184 -0.184
brwsrFr:DDG  0.267 -0.725 -0.186 -0.369 -0.189 -0.189 -0.189  0.510
brwsrFrfx:G  0.264 -0.718 -0.184 -0.190 -0.373 -0.187 -0.187  0.505  0.521
brwsrFrf:Y!  0.260 -0.707 -0.181 -0.187 -0.184 -0.368 -0.184  0.497  0.513  0.509
brwsrFrfx:Y  0.262 -0.711 -0.182 -0.188 -0.185 -0.185 -0.373  0.500  0.517  0.511  0.503
```

## Response: RBO (p=0.95)

```
Linear mixed model fit by REML. t-tests use Satterthwaite's method ['lmerModLmerTest']
Formula: rbo_95 ~ browser * engine + (1 | agent1) + (1 | agent2) + (1 |      machine_combination)
   Data: subdf

REML criterion at convergence: -5195.3
```

```
Scaled residuals:
    Min      1Q  Median      3Q     Max
-3.9502 -0.3593 -0.0084  0.2876  7.3957

Random effects:
 Groups              Name        Variance Std.Dev.
 machine_combination (Intercept) 0.002398 0.04897
 agent2              (Intercept) 0.001885 0.04342
 agent1              (Intercept) 0.008847 0.09406
 Residual                        0.002217 0.04708
Number of obs: 2258, groups:  machine_combination, 1537; agent2, 192; agent1, 96

Fixed effects:
                           Estimate Std. Error         df t value Pr(>|t|)
(Intercept)                0.865571   0.026693 138.305702  32.427  < 2e-16 ***
browserFirefox            -0.001450   0.017449 170.485230  -0.083 0.933893
engineBing                -0.109011   0.038411 138.838937  -2.838 0.005220 **
engineDDG                 -0.274974   0.037146 137.665772  -7.403  1.2e-11 ***
engineGoogle              -0.034586   0.037733 138.087898  -0.917 0.360957
engineYahoo!              -0.083507   0.037748 138.289345  -2.212 0.028593 *
engineYandex              -0.065342   0.037779 138.698183  -1.730 0.085929 .
browserFirefox:engineBing  0.020832   0.024413 163.493650   0.853 0.394741
browserFirefox:engineDDG   0.095367   0.024047 168.765669   3.966 0.000108 ***
browserFirefox:engineGoogle 0.003338  0.024305 170.943935   0.137 0.890921
browserFirefox:engineYahoo! -0.017142 0.024673 170.437729  -0.695 0.488134
browserFirefox:engineYandex -0.003268 0.024536 171.496064  -0.133 0.894196
---
Signif. codes:  0 '***' 0.001 '**' 0.01 '*' 0.05 '.' 0.1 ' ' 1

Correlation of Fixed Effects:
            (Intr) brwsrF engnBn engDDG engnGg engnY! engnYn brwF:B bF:DDG brwF:G brF:Y!
browserFrfx -0.343
engineBing  -0.695  0.238
engineDDG   -0.719  0.246  0.499
engineGoogl -0.707  0.242  0.492  0.508
engineYaho! -0.707  0.242  0.491  0.508  0.500
engineYandx -0.707  0.242  0.491  0.508  0.500  0.500
brwsrFrfx:B  0.245 -0.693 -0.355 -0.176 -0.173 -0.173 -0.173
brwsrFr:DDG  0.249 -0.726 -0.173 -0.343 -0.176 -0.176 -0.176  0.503
brwsrFrfx:G  0.246 -0.718 -0.171 -0.177 -0.347 -0.174 -0.174  0.498  0.521
brwsrFrf:Y!  0.242 -0.707 -0.168 -0.174 -0.171 -0.343 -0.171  0.490  0.513  0.522
brwsrFrfx:Y  0.244 -0.711 -0.169 -0.175 -0.172 -0.172 -0.347  0.493  0.530  0.511  0.503
```

## Response: RBO (p=0.8)

```
Linear mixed model fit by REML. t-tests use Satterthwaite's method ['lmerModLmerTest']
Formula: rbo_80 ~ browser * engine + (1 | agent1) + (1 | agent2) + (1 |      machine_combination)
   Data: subdf

REML criterion at convergence: -2334.4

Scaled residuals:
    Min      1Q  Median      3Q     Max
-4.3219 -0.5208  0.0167  0.2249  6.3630

Random effects:
 Groups              Name        Variance  Std.Dev.
 machine_combination (Intercept) 0.0003591 0.01895
```

```
 agent2          (Intercept) 0.0032672 0.05716
 agent1          (Intercept) 0.0110980 0.10535
 Residual                    0.0160351 0.12663
Number of obs: 2258, groups:  machine_combination, 1537; agent2, 192; agent1, 96

Fixed effects:
                          Estimate Std. Error       df t value Pr(>|t|)
(Intercept)               0.981113   0.032308 166.961350  30.368  < 2e-16 ***
browserFirefox            0.003848   0.025225 202.508689   0.153  0.87891
engineBing               -0.060564   0.046553 168.514312  -1.301  0.19505
engineDDG                -0.539083   0.044896 165.330342 -12.007  < 2e-16 ***
engineGoogle             -0.132697   0.045655 166.552255  -2.907  0.00415 **
engineYahoo!             -0.016492   0.045689 166.952364  -0.361  0.71858
engineYandex             -0.144938   0.045754 167.718450  -3.168  0.00183 **
browserFirefox:engineBing   0.008107   0.035911 205.149901   0.226  0.82162
browserFirefox:engineDDG    0.076198   0.034702 199.413537   2.196  0.02926 *
browserFirefox:engineGoogle -0.021174   0.035153 203.712744  -0.602  0.54761
browserFirefox:engineYahoo! -0.077796   0.035672 202.573259  -2.181  0.03034 *
browserFirefox:engineYandex -0.010141   0.035505 203.951526  -0.286  0.77546
---
Signif. codes:  0 '***' 0.001 '**' 0.01 '*' 0.05 '.' 0.1 ' ' 1

Correlation of Fixed Effects:
            (Intr) brwsrF engnBn engDDG engnGg engnY! engnYn brwF:B bF:DDG brwF:G brF:Y!
browserFrfx -0.430
engineBing  -0.694  0.298
engineDDG   -0.720  0.309  0.499
engineGoogl -0.708  0.304  0.491  0.509
engineYaho! -0.707  0.304  0.491  0.509  0.500
engineYandx -0.706  0.303  0.490  0.508  0.500  0.499
brwsrFrfx:B  0.302 -0.701 -0.439 -0.217 -0.214 -0.213 -0.213
brwsrFr:DDG  0.312 -0.727 -0.217 -0.430 -0.221 -0.221 -0.221  0.509
brwsrFrfx:G  0.308 -0.718 -0.214 -0.222 -0.434 -0.218 -0.218  0.503  0.522
brwsrFrf:Y!  0.304 -0.707 -0.211 -0.219 -0.215 -0.430 -0.215  0.496  0.514  0.508
brwsrFrfx:Y  0.305 -0.710 -0.212 -0.220 -0.216 -0.216 -0.435  0.498  0.517  0.510  0.502
convergence code: 0
Model failed to converge with max|grad| = 0.00267193 (tol = 0.002, component 1)
```

## B.2 Comparison of Search Engines for "bernie sanders" query

### Response:  Jaccard

```
Linear mixed model fit by REML. t-tests use Satterthwaite's method ['lmerModLmerTest']
Formula: jaccard ~ engines_ + (1 | agent1) + (1 | agent2) + (1 | browser) +      (1 | machine_combination)
   Data: subdf

REML criterion at convergence: -38073.9

Scaled residuals:
     Min      1Q  Median      3Q     Max
-13.8010 -0.2851 -0.0285  0.2514  5.0431

Random effects:
 Groups              Name        Variance  Std.Dev.
 machine_combination (Intercept) 0.000e+00 0.000000
```

```
agent2              (Intercept) 8.216e-05 0.009064
agent1              (Intercept) 5.778e-05 0.007602
browser             (Intercept) 0.000e+00 0.000000
Residual                        3.617e-04 0.019018
Number of obs: 7677, groups:  machine_combination, 4074; agent2, 192; agent1, 192; browser, 2

Fixed effects:
                       Estimate Std. Error         df t value Pr(>|t|)
(Intercept)           8.901e-03  1.728e-03  5.222e+02   5.153 3.65e-07 ***
engines_Baidu-DDG     2.172e-02  1.888e-03  7.988e+02  11.508  < 2e-16 ***
engines_Baidu-Google  1.334e-02  1.902e-03  7.991e+02   7.013 4.98e-12 ***
engines_Baidu-Yahoo!  1.867e-02  1.932e-03  8.005e+02   9.665  < 2e-16 ***
engines_Baidu-Yandex  5.711e-02  1.920e-03  8.055e+02  29.737  < 2e-16 ***
engines_Bing-DDG      1.150e-01  1.888e-03  8.022e+02  60.904  < 2e-16 ***
engines_Bing-Google   1.332e-01  1.903e-03  8.020e+02  69.999  < 2e-16 ***
engines_Bing-Yahoo!   1.093e-01  1.933e-03  8.028e+02  56.540  < 2e-16 ***
engines_Bing-Yandex   8.261e-02  1.918e-03  8.022e+02  43.074  < 2e-16 ***
engines_DDG-Google    4.069e-02  2.378e-03  5.022e+02  17.112  < 2e-16 ***
engines_DDG-Yahoo!    5.783e-01  2.400e-03  5.053e+02 240.975  < 2e-16 ***
engines_DDG-Yandex    8.400e-02  2.388e-03  5.034e+02  35.173  < 2e-16 ***
engines_Google-Yahoo! 5.740e-02  2.412e-03  5.080e+02  23.797  < 2e-16 ***
engines_Google-Yandex 1.060e-01  2.400e-03  5.059e+02  44.175  < 2e-16 ***
engines_Yahoo!-Yandex 9.865e-02  2.424e-03  5.105e+02  40.698  < 2e-16 ***
---
Signif. codes:  0 '***' 0.001 '**' 0.01 '*' 0.05 '.' 0.1 ' ' 1
convergence code: 0
boundary (singular) fit: see ?isSingular
```

## Response: Jaccard Top-10

```
Linear mixed model fit by REML. t-tests use Satterthwaite's method ['lmerModLmerTest']
Formula: jacctop10 ~ engines_ + (1 | agent1) + (1 | agent2) + (1 | browser) +      (1 | machine_combination)
   Data: subdf

REML criterion at convergence: -29867.6

Scaled residuals:
    Min      1Q  Median      3Q     Max
-4.9982 -0.3460 -0.0364  0.3465  7.8219

Random effects:
 Groups              Name        Variance  Std.Dev.
 machine_combination (Intercept) 0.0000145 0.003808
 agent2              (Intercept) 0.0006027 0.024551
 agent1              (Intercept) 0.0006692 0.025869
 browser             (Intercept) 0.0001022 0.010108
 Residual                        0.0009846 0.031378
Number of obs: 7677, groups:  machine_combination, 4074; agent2, 192; agent1, 192; browser, 2

Fixed effects:
                       Estimate Std. Error         df t value Pr(>|t|)
(Intercept)           2.306e-04  8.586e-03  1.395e+00   0.027    0.982
engines_Baidu-DDG     6.607e-03  4.863e-03  5.027e+02   1.359    0.175
engines_Baidu-Google -2.886e-04  4.900e-03  5.035e+02  -0.059    0.953
engines_Baidu-Yahoo!  4.315e-02  4.976e-03  5.032e+02   8.671   <2e-16 ***
engines_Baidu-Yandex  4.723e-02  4.940e-03  5.048e+02   9.561   <2e-16 ***
engines_Bing-DDG      9.854e-02  4.862e-03  5.024e+02  20.265   <2e-16 ***
```

```
engines_Bing-Google     3.388e-01  4.900e-03  5.033e+02  69.148  <2e-16 ***
engines_Bing-Yahoo!     2.035e-01  4.976e-03  5.034e+02  40.894  <2e-16 ***
engines_Bing-Yandex     2.598e-01  4.937e-03  5.036e+02  52.630  <2e-16 ***
engines_DDG-Google      1.241e-01  6.586e-03  4.174e+02  18.847  <2e-16 ***
engines_DDG-Yahoo!      2.475e-01  6.641e-03  4.183e+02  37.261  <2e-16 ***
engines_DDG-Yandex      1.498e-01  6.614e-03  4.181e+02  22.652  <2e-16 ***
engines_Google-Yahoo!   2.067e-01  6.670e-03  4.196e+02  30.983  <2e-16 ***
engines_Google-Yandex   1.133e-01  6.639e-03  4.186e+02  17.071  <2e-16 ***
engines_Yahoo!-Yandex   1.914e-01  6.696e-03  4.200e+02  28.580  <2e-16 ***
---
Signif. codes:  0 '***' 0.001 '**' 0.01 '*' 0.05 '.' 0.1 ' ' 1
convergence code: 0
Model failed to converge with max|grad| = 0.106145 (tol = 0.002, component 1)
```

## Response: RBO (p=0.95)

```
Linear mixed model fit by REML. t-tests use Satterthwaite's method ['lmerModLmerTest']
Formula: rbo_95 ~ engines_ + (1 | agent1) + (1 | agent2) + (1 | browser) +      (1 | machine_combination)
   Data: subdf

REML criterion at convergence: -34727.5

Scaled residuals:
    Min      1Q  Median      3Q     Max
-4.6231 -0.4584 -0.0280  0.4810  4.6119

Random effects:
 Groups              Name        Variance  Std.Dev.
 machine_combination (Intercept) 0.0000000 0.00000
 agent2              (Intercept) 0.0003933 0.01983
 agent1              (Intercept) 0.0004151 0.02037
 browser             (Intercept) 0.0001345 0.01160
 Residual                        0.0005250 0.02291
Number of obs: 7677, groups:  machine_combination, 4074; agent2, 192; agent1, 192; browser, 2

Fixed effects:
                       Estimate Std. Error         df t value Pr(>|t|)
(Intercept)           6.244e-03  9.023e-03  1.299e+00   0.692   0.5916
engines_Baidu-DDG     2.634e-02  3.820e-03  4.797e+02   6.896  1.7e-11 ***
engines_Baidu-Google  9.471e-03  3.848e-03  4.803e+02   2.461   0.0142 *
engines_Baidu-Yahoo!  4.255e-02  3.908e-03  4.802e+02  10.888  < 2e-16 ***
engines_Baidu-Yandex  6.158e-02  3.880e-03  4.815e+02  15.872  < 2e-16 ***
engines_Bing-DDG      1.375e-01  3.819e-03  4.795e+02  36.004  < 2e-16 ***
engines_Bing-Google   3.381e-01  3.849e-03  4.803e+02  87.847  < 2e-16 ***
engines_Bing-Yahoo!   2.000e-01  3.908e-03  4.804e+02  51.165  < 2e-16 ***
engines_Bing-Yandex   2.211e-01  3.878e-03  4.805e+02  57.011  < 2e-16 ***
engines_DDG-Google    1.396e-01  5.212e-03  4.102e+02  26.788  < 2e-16 ***
engines_DDG-Yahoo!    3.717e-01  5.255e-03  4.109e+02  70.724  < 2e-16 ***
engines_DDG-Yandex    2.057e-01  5.233e-03  4.105e+02  39.313  < 2e-16 ***
engines_Google-Yahoo! 2.052e-01  5.276e-03  4.117e+02  38.892  < 2e-16 ***
engines_Google-Yandex 2.033e-01  5.254e-03  4.112e+02  38.707  < 2e-16 ***
engines_Yahoo!-Yandex 2.378e-01  5.298e-03  4.123e+02  44.887  < 2e-16 ***
---
Signif. codes:  0 '***' 0.001 '**' 0.01 '*' 0.05 '.' 0.1 ' ' 1
convergence code: 0
boundary (singular) fit: see ?isSingular
```

## Response: RBO (p=0.8)

```
Linear mixed model fit by REML. t-tests use Satterthwaite's method ['lmerModLmerTest']
Formula: rbo_80 ~ engines_ + (1 | agent1) + (1 | agent2) + (1 | browser) +      (1 | machine_combination)
   Data: subdf

REML criterion at convergence: -21470.8

Scaled residuals:
    Min      1Q  Median      3Q     Max
-3.2779 -0.5267 -0.0294  0.4932  4.8220

Random effects:
 Groups              Name        Variance  Std.Dev.
 machine_combination (Intercept) 4.670e-06 0.002161
 agent2              (Intercept) 1.947e-03 0.044120
 agent1              (Intercept) 1.918e-03 0.043790
 browser             (Intercept) 8.221e-04 0.028672
 Residual                        2.981e-03 0.054602
Number of obs: 7677, groups:  machine_combination, 4074; agent2, 192; agent1, 192; browser, 2

Fixed effects:
                      Estimate Std. Error         df t value Pr(>|t|)
(Intercept)           3.294e-04  2.190e-02  1.238e+00   0.015   0.9901
engines_Baidu-DDG     3.727e-03  8.469e-03  5.003e+02   0.440   0.6600
engines_Baidu-Google  2.906e-04  8.532e-03  5.009e+02   0.034   0.9728
engines_Baidu-Yahoo!  1.294e-02  8.665e-03  5.008e+02   1.493   0.1361
engines_Baidu-Yandex  1.583e-02  8.603e-03  5.024e+02   1.839   0.0664 .
engines_Bing-DDG      1.037e-01  8.468e-03  5.002e+02  12.241   <2e-16 ***
engines_Bing-Google   3.557e-01  8.533e-03  5.010e+02  41.681   <2e-16 ***
engines_Bing-Yahoo!   1.793e-01  8.665e-03  5.011e+02  20.696   <2e-16 ***
engines_Bing-Yandex   1.995e-01  8.598e-03  5.012e+02  23.207   <2e-16 ***
engines_DDG-Google    1.311e-01  1.148e-02  4.172e+02  11.416   <2e-16 ***
engines_DDG-Yahoo!    2.895e-01  1.158e-02  4.181e+02  25.009   <2e-16 ***
engines_DDG-Yandex    2.062e-01  1.153e-02  4.174e+02  17.886   <2e-16 ***
engines_Google-Yahoo! 2.809e-01  1.162e-02  4.188e+02  24.169   <2e-16 ***
engines_Google-Yandex 2.484e-01  1.157e-02  4.183e+02  21.466   <2e-16 ***
engines_Yahoo!-Yandex 3.109e-01  1.167e-02  4.197e+02  26.633   <2e-16 ***
---
Signif. codes:  0 '***' 0.001 '**' 0.01 '*' 0.05 '.' 0.1 ' ' 1
convergence code: 0
Model failed to converge with max|grad| = 0.00230499 (tol = 0.002, component 1)
```

# C. Statistical Tests for "donald trump" query

## C.1 Comparison of Browsers for "donald trump" query

### Response: Jaccard

```
Linear mixed model fit by REML. t-tests use Satterthwaite's method ['lmerModLmerTest']
Formula: jaccard ~ browser * engine + (1 | agent1) + (1 | agent2) + (1 |     machine_combination)
```

Data: subdf

REML criterion at convergence: -6854.8

Scaled residuals:
     Min      1Q  Median      3Q     Max
 -2.9690 -0.4603 -0.0063  0.2848 13.3648

Random effects:
 Groups              Name        Variance  Std.Dev.
 machine_combination (Intercept) 2.402e-05 0.004901
 agent2              (Intercept) 3.030e-03 0.055044
 agent1              (Intercept) 3.489e-03 0.059071
 Residual                        1.802e-03 0.042447
Number of obs: 2258, groups:  machine_combination, 1537; agent2, 192; agent1, 96

Fixed effects:
                             Estimate Std. Error        df t value Pr(>|t|)
(Intercept)                 9.268e-01  2.060e-02 2.302e+02  44.996  < 2e-16 ***
browserFirefox              1.780e-02  2.039e-02 1.790e+02   0.873   0.3839
engineBing                 -1.358e-01  2.963e-02 2.307e+02  -4.583  7.5e-06 ***
engineDDG                   4.789e-02  2.868e-02 2.296e+02   1.670   0.0963 .
engineGoogle               -2.340e-02  2.911e-02 2.297e+02  -0.804   0.4223
engineYahoo!                6.956e-02  2.912e-02 2.301e+02   2.388   0.0177 *
engineYandex               -2.983e-02  2.912e-02 2.298e+02  -1.025   0.3066
browserFirefox:engineBing  -1.308e-04  2.887e-02 1.798e+02  -0.005   0.9964
browserFirefox:engineDDG   -2.224e-02  2.816e-02 1.785e+02  -0.790   0.4308
browserFirefox:engineGoogle -5.367e-03 2.839e-02 1.789e+02  -0.189   0.8503
browserFirefox:engineYahoo! -4.614e-02 2.883e-02 1.789e+02  -1.600   0.1113
browserFirefox:engineYandex -1.665e-02 2.860e-02 1.789e+02  -0.582   0.5612
---
Signif. codes:  0 '***' 0.001 '**' 0.01 '*' 0.05 '.' 0.1 ' ' 1

Correlation of Fixed Effects:
            (Intr) brwsrF engnBn engDDG engnGg engnY! engnYn brwF:B bF:DDG brwF:G brF:Y!
browserFrfx -0.491
engineBing  -0.695  0.341
engineDDG   -0.718  0.353  0.499
engineGoogl -0.708  0.347  0.492  0.508
engineYaho! -0.707  0.347  0.492  0.508  0.500
engineYandx -0.707  0.347  0.492  0.508  0.500  0.500
brwsrFrfx:B  0.347 -0.706 -0.499 -0.249 -0.245 -0.245 -0.245
brwsrFr:DDG  0.355 -0.724 -0.247 -0.494 -0.251 -0.251 -0.251  0.511
brwsrFrfx:G  0.353 -0.718 -0.245 -0.253 -0.498 -0.249 -0.249  0.507  0.520
brwsrFrf:Y!  0.347 -0.707 -0.241 -0.249 -0.246 -0.491 -0.246  0.499  0.512  0.508
brwsrFrfx:Y  0.350 -0.713 -0.243 -0.251 -0.248 -0.248 -0.494  0.504  0.516  0.512  0.504

## Response: Jaccard Top-10

Linear mixed model fit by REML. t-tests use Satterthwaite's method ['lmerModLmerTest']
Formula: jacctop10 ~ browser * engine + (1 | agent1) + (1 | agent2) +      (1 | machine_combination)
   Data: subdf

REML criterion at convergence: -2583

Scaled residuals:
     Min      1Q  Median      3Q     Max
 -4.1699 -0.4710  0.0207  0.2965  7.9511

```
Random effects:
 Groups             Name        Variance Std.Dev.
 machine_combination (Intercept) 0.000000 0.00000
 agent2             (Intercept) 0.007879 0.08877
 agent1             (Intercept) 0.020055 0.14162
 Residual                       0.013301 0.11533
Number of obs: 2258, groups:  machine_combination, 1537; agent2, 192; agent1, 96

Fixed effects:
                             Estimate Std. Error       df t value Pr(>|t|)
(Intercept)                   0.87962    0.04320 169.94158  20.361  < 2e-16 ***
browserFirefox                0.06117    0.03455 172.64433   1.770   0.0784 .
engineBing                   -0.05022    0.06217 170.65705  -0.808   0.4203
engineDDG                    -0.41453    0.06013 169.28102  -6.894 1.03e-10 ***
engineGoogle                 -0.11447    0.06105 169.45021  -1.875   0.0625 .
engineYahoo!                  0.12038    0.06109 169.87225   1.971   0.0504 .
engineYandex                 -0.09304    0.06106 169.63212  -1.524   0.1295
browserFirefox:engineBing    -0.08163    0.04902 175.08306  -1.665   0.0977 .
browserFirefox:engineDDG      0.03586    0.04767 171.52404   0.752   0.4530
browserFirefox:engineGoogle  -0.01240    0.04809 172.62547  -0.258   0.7968
browserFirefox:engineYahoo!  -0.08617    0.04885 172.65236  -1.764   0.0795 .
browserFirefox:engineYandex  -0.06200    0.04845 172.59661  -1.280   0.2023
---
Signif. codes:  0 '***' 0.001 '**' 0.01 '*' 0.05 '.' 0.1 ' ' 1

Correlation of Fixed Effects:
            (Intr) brwsrF engnBn engDDG engnGg engnY! engnYn brwF:B bF:DDG brwF:G brF:Y!
browserFrfx -0.411
engineBing  -0.695  0.285
engineDDG   -0.718  0.295  0.499
engineGoogl -0.708  0.291  0.492  0.508
engineYaho! -0.707  0.290  0.491  0.508  0.500
engineYandx -0.707  0.291  0.492  0.508  0.501  0.500
brwsrFrfx:B  0.289 -0.705 -0.418 -0.208 -0.205 -0.205 -0.205
brwsrFr:DDG  0.298 -0.725 -0.207 -0.413 -0.211 -0.210 -0.211  0.511
brwsrFrfx:G  0.295 -0.718 -0.205 -0.212 -0.415 -0.209 -0.209  0.506  0.521
brwsrFrf:Y!  0.290 -0.707 -0.202 -0.209 -0.206 -0.410 -0.205  0.498  0.513  0.508
brwsrFrfx:Y  0.293 -0.713 -0.203 -0.210 -0.207 -0.207 -0.413  0.503  0.517  0.512  0.504
convergence code: 0
boundary (singular) fit: see ?isSingular
```

## Response:  RBO (p=0.95)

```
Linear mixed model fit by REML. t-tests use Satterthwaite's method ['lmerModLmerTest']
Formula: rbo_95 ~ browser * engine + (1 | agent1) + (1 | agent2) + (1 |      machine_combination)
   Data: subdf

REML criterion at convergence: -4790.1

Scaled residuals:
    Min      1Q  Median      3Q     Max
-3.5045 -0.3899 -0.0018  0.2495  6.7850

Random effects:
 Groups             Name        Variance Std.Dev.
 machine_combination (Intercept) 0.001331 0.03649
```

```
 agent2           (Intercept) 0.002855 0.05343
 agent1           (Intercept) 0.007969 0.08927
 Residual                     0.003759 0.06131
Number of obs: 2258, groups:  machine_combination, 1537; agent2, 192; agent1, 96

Fixed effects:
                           Estimate Std. Error       df t value Pr(>|t|)
(Intercept)                 0.80494    0.02688 163.74001  29.946  < 2e-16 ***
browserFirefox              0.04351    0.02088 168.40027   2.084   0.0387 *
engineBing                 -0.08110    0.03868 164.41988  -2.097   0.0376 *
engineDDG                  -0.18151    0.03741 163.11546  -4.851 2.85e-06 ***
engineGoogle               -0.05988    0.03798 163.27919  -1.576   0.1169
engineYahoo!               -0.06219    0.03801 163.67630  -1.636   0.1037
engineYandex               -0.03442    0.03800 163.44998  -0.906   0.3663
browserFirefox:engineBing  -0.04529    0.02946 167.33422  -1.538   0.1260
browserFirefox:engineDDG    0.01359    0.02881 167.27650   0.472   0.6378
browserFirefox:engineGoogle -0.04003    0.02907 168.40620  -1.377   0.1703
browserFirefox:engineYahoo! -0.05773    0.02952 168.41115  -1.955   0.0522 .
browserFirefox:engineYandex -0.03782    0.02928 168.36001  -1.292   0.1983
---
Signif. codes:  0 '***' 0.001 '**' 0.01 '*' 0.05 '.' 0.1 ' ' 1

Correlation of Fixed Effects:
            (Intr) brwsrF engnBn engDDG engnGg engnY! engnYn brwF:B bF:DDG brwF:G brF:Y!
browserFrfx -0.400
engineBing  -0.695  0.278
engineDDG   -0.718  0.287  0.499
engineGoogl -0.708  0.283  0.492  0.508
engineYaho! -0.707  0.283  0.491  0.508  0.500
engineYandx -0.707  0.283  0.492  0.508  0.501  0.500
brwsrFrfx:B  0.284 -0.700 -0.410 -0.204 -0.201 -0.201 -0.201
brwsrFr:DDG  0.290 -0.725 -0.201 -0.402 -0.205 -0.205 -0.205  0.508
brwsrFrfx:G  0.287 -0.718 -0.200 -0.206 -0.405 -0.203 -0.203  0.503  0.521
brwsrFrf:Y!  0.283 -0.707 -0.197 -0.203 -0.200 -0.400 -0.200  0.495  0.513  0.514
brwsrFrfx:Y  0.285 -0.713 -0.198 -0.205 -0.202 -0.202 -0.402  0.500  0.522  0.512  0.504
```

## Response: RBO (p=0.8)

Linear mixed model fit by REML. t-tests use Satterthwaite's method ['lmerModLmerTest']
Formula: rbo_80 ~ browser * engine + (1 | agent1) + (1 | agent2) + (1 |      machine_combination)
   Data: subdf

REML criterion at convergence: -1917.5

```
Scaled residuals:
    Min      1Q  Median      3Q     Max
-3.7763 -0.5061  0.0169  0.3004  6.8509

Random effects:
 Groups              Name        Variance  Std.Dev.
 machine_combination (Intercept) 1.002e-10 1.001e-05
 agent2              (Intercept) 4.265e-03 6.531e-02
 agent1              (Intercept) 1.695e-02 1.302e-01
 Residual                        1.944e-02 1.394e-01
Number of obs: 2258, groups:  machine_combination, 1537; agent2, 192; agent1, 96

Fixed effects:
                           Estimate Std. Error       df t value Pr(>|t|)
```

```
(Intercept)                 8.703e-01  3.871e-02  1.515e+02  22.485  < 2e-16 ***
browserFirefox              6.573e-02  2.836e-02  1.545e+02   2.318  0.02179 *
engineBing                 -4.334e-02  5.576e-02  1.527e+02  -0.777  0.43820
engineDDG                  -4.560e-01  5.384e-02  1.505e+02  -8.470 2.06e-14 ***
engineGoogle               -1.829e-01  5.468e-02  1.509e+02  -3.344  0.00104 **
engineYahoo!                3.164e-02  5.474e-02  1.515e+02   0.578  0.56415
engineYandex               -5.890e-02  5.471e-02  1.512e+02  -1.077  0.28331
browserFirefox:engineBing  -7.082e-02  4.041e-02  1.594e+02  -1.753  0.08158 .
browserFirefox:engineDDG    4.438e-04  3.906e-02  1.524e+02   0.011  0.99095
browserFirefox:engineGoogle -5.831e-02  3.949e-02  1.551e+02  -1.477  0.14180
browserFirefox:engineYahoo! -6.964e-02  4.011e-02  1.548e+02  -1.736  0.08450 .
browserFirefox:engineYandex -6.077e-02  3.978e-02  1.548e+02  -1.528  0.12861
---
Signif. codes:  0 '***' 0.001 '**' 0.01 '*' 0.05 '.' 0.1 ' ' 1

Correlation of Fixed Effects:
            (Intr) brwsrF engnBn engDDG engnGg engnY! engnYn brwF:B bF:DDG brwF:G brF:Y!
browserFrfx -0.400
engineBing  -0.694  0.278
engineDDG   -0.719  0.288  0.499
engineGoogl -0.708  0.283  0.491  0.509
engineYaho! -0.707  0.283  0.491  0.508  0.501
engineYandx -0.708  0.283  0.491  0.509  0.501  0.500
brwsrFrfx:B  0.281 -0.702 -0.408 -0.202 -0.199 -0.198 -0.199
brwsrFr:DDG  0.290 -0.726 -0.202 -0.401 -0.206 -0.205 -0.205  0.510
brwsrFrfx:G  0.287 -0.718 -0.199 -0.206 -0.404 -0.203 -0.203  0.504  0.521
brwsrFrf:Y!  0.283 -0.707 -0.196 -0.203 -0.200 -0.400 -0.200  0.496  0.513  0.508
brwsrFrfx:Y  0.285 -0.713 -0.198 -0.205 -0.202 -0.202 -0.402  0.500  0.518  0.512  0.504
convergence code: 0
boundary (singular) fit: see ?isSingular
```

## C.2 Comparison of Search Engines for "donald trump" query

### Response:  Jaccard

```
Linear mixed model fit by REML. t-tests use Satterthwaite's method ['lmerModLmerTest']
Formula: jaccard ~ engines_ + (1 | agent1) + (1 | agent2) + (1 | browser) +      (1 | machine_combination)
   Data: subdf

REML criterion at convergence: -44721.3

Scaled residuals:
    Min      1Q  Median      3Q     Max
-9.2712 -0.3560 -0.0044  0.3729  5.8130

Random effects:
 Groups              Name        Variance  Std.Dev.
 machine_combination (Intercept) 6.827e-06 0.002613
 agent2              (Intercept) 3.995e-05 0.006321
 agent1              (Intercept) 4.211e-05 0.006489
 browser             (Intercept) 1.190e-06 0.001091
 Residual                        1.428e-04 0.011949
Number of obs: 7677, groups:  machine_combination, 4074; agent2, 192; agent1, 192; browser, 2

Fixed effects:
```

```
                        Estimate Std. Error        df t value Pr(>|t|)
(Intercept)            1.141e-02  1.500e-03 4.659e+00   7.608 0.000849 ***
engines_Baidu-DDG      7.656e-02  1.373e-03 6.884e+02  55.744  < 2e-16 ***
engines_Baidu-Google   5.355e-02  1.383e-03 6.881e+02  38.716  < 2e-16 ***
engines_Baidu-Yahoo!   8.796e-02  1.404e-03 6.878e+02  62.628  < 2e-16 ***
engines_Baidu-Yandex   4.834e-02  1.394e-03 6.898e+02  34.668  < 2e-16 ***
engines_Bing-DDG       1.473e-02  1.374e-03 6.888e+02  10.721  < 2e-16 ***
engines_Bing-Google    1.587e-01  1.384e-03 6.898e+02 114.618  < 2e-16 ***
engines_Bing-Yahoo!    4.115e-03  1.404e-03 6.876e+02   2.930 0.003506 **
engines_Bing-Yandex    4.258e-04  1.393e-03 6.877e+02   0.306 0.760051
engines_DDG-Google     1.390e-01  1.776e-03 4.782e+02  78.255  < 2e-16 ***
engines_DDG-Yahoo!     5.576e-01  1.792e-03 4.807e+02 311.095  < 2e-16 ***
engines_DDG-Yandex     1.919e-01  1.787e-03 4.824e+02 107.373  < 2e-16 ***
engines_Google-Yahoo!  1.235e-01  1.803e-03 4.843e+02  68.510  < 2e-16 ***
engines_Google-Yandex  1.062e-01  1.791e-03 4.802e+02  59.320  < 2e-16 ***
engines_Yahoo!-Yandex  1.618e-01  1.808e-03 4.832e+02  89.491  < 2e-16 ***
---
Signif. codes:  0 '***' 0.001 '**' 0.01 '*' 0.05 '.' 0.1 ' ' 1
```

## Response: Jaccard Top-10

```
Linear mixed model fit by REML. t-tests use Satterthwaite's method ['lmerModLmerTest']
Formula: jacctop10 ~ engines_ + (1 | agent1) + (1 | agent2) + (1 | browser) +      (1 | machine_combination)
   Data: subdf

REML criterion at convergence: -33106.1

Scaled residuals:
    Min      1Q  Median      3Q     Max
-7.2345 -0.4500 -0.0214  0.3627  5.1511

Random effects:
 Groups              Name        Variance  Std.Dev.
 machine_combination (Intercept) 4.422e-12 2.103e-06
 agent2              (Intercept) 1.195e-04 1.093e-02
 agent1              (Intercept) 1.738e-04 1.318e-02
 browser             (Intercept) 4.091e-07 6.396e-04
 Residual                        6.889e-04 2.625e-02
Number of obs: 7677, groups:  machine_combination, 4074; agent2, 192; agent1, 192; browser, 2

Fixed effects:
                        Estimate Std. Error        df t value Pr(>|t|)
(Intercept)            4.635e-02  2.511e-03 2.565e+01  18.456 2.51e-16 ***
engines_Baidu-DDG      5.389e-02  2.681e-03 7.670e+02  20.101  < 2e-16 ***
engines_Baidu-Google  -4.039e-02  2.700e-03 7.659e+02 -14.960  < 2e-16 ***
engines_Baidu-Yahoo!   7.721e-02  2.741e-03 7.650e+02  28.166  < 2e-16 ***
engines_Baidu-Yandex  -4.613e-02  2.723e-03 7.691e+02 -16.943  < 2e-16 ***
engines_Bing-DDG       2.493e-04  2.680e-03 7.631e+02   0.093    0.926
engines_Bing-Google    2.398e-01  2.701e-03 7.646e+02  88.784  < 2e-16 ***
engines_Bing-Yahoo!    1.247e-02  2.740e-03 7.598e+02   4.553 6.17e-06 ***
engines_Bing-Yandex   -4.272e-02  2.718e-03 7.604e+02 -15.717  < 2e-16 ***
engines_DDG-Google    -4.129e-02  3.400e-03 4.926e+02 -12.144  < 2e-16 ***
engines_DDG-Yahoo!     3.545e-01  3.432e-03 4.960e+02 103.272  < 2e-16 ***
engines_DDG-Yandex     3.071e-02  3.416e-03 4.944e+02   8.989  < 2e-16 ***
engines_Google-Yahoo! -3.902e-02  3.446e-03 4.969e+02 -11.323  < 2e-16 ***
engines_Google-Yandex -4.642e-02  3.429e-03 4.954e+02 -13.535  < 2e-16 ***
engines_Yahoo!-Yandex  4.386e-02  3.463e-03 4.992e+02  12.664  < 2e-16 ***
---
```

Signif. codes:  0 '***' 0.001 '**' 0.01 '*' 0.05 '.' 0.1 ' ' 1
convergence code: 0
boundary (singular) fit: see ?isSingular

## Response:  RBO (p=0.95)

Linear mixed model fit by REML. t-tests use Satterthwaite's method ['lmerModLmerTest']
Formula: rbo_95 ~ engines_ + (1 | agent1) + (1 | agent2) + (1 | browser) +      (1 | machine_combination)
   Data: subdf

REML criterion at convergence: -38490.9

Scaled residuals:
    Min      1Q  Median      3Q     Max
-6.8115 -0.3709  0.0242  0.4396  4.9614

Random effects:
 Groups              Name        Variance  Std.Dev.
 machine_combination (Intercept) 6.322e-06 0.002514
 agent2              (Intercept) 1.511e-04 0.012293
 agent1              (Intercept) 1.712e-04 0.013084
 browser             (Intercept) 8.280e-07 0.000910
 Residual                        3.220e-04 0.017944
Number of obs: 7677, groups:  machine_combination, 4074; agent2, 192; agent1, 192; browser, 2

Fixed effects:
                       Estimate Std. Error         df t value Pr(>|t|)
(Intercept)            0.068077   0.002513 426.304599  27.093  < 2e-16 ***
engines_Baidu-DDG      0.063290   0.002510 547.663603  25.216  < 2e-16 ***
engines_Baidu-Google  -0.034456   0.002528 547.468362 -13.631  < 2e-16 ***
engines_Baidu-Yahoo!   0.097339   0.002567 547.294878  37.921  < 2e-16 ***
engines_Baidu-Yandex  -0.043600   0.002548 548.585723 -17.110  < 2e-16 ***
engines_Bing-DDG       0.009077   0.002510 547.587729   3.616 0.000327 ***
engines_Bing-Google    0.307000   0.002529 548.212508 121.383  < 2e-16 ***
engines_Bing-Yahoo!    0.023445   0.002567 546.855344   9.134  < 2e-16 ***
engines_Bing-Yandex   -0.058837   0.002547 546.925811 -23.103  < 2e-16 ***
engines_DDG-Google    -0.022523   0.003360 434.515188  -6.703 6.34e-11 ***
engines_DDG-Yahoo!     0.375583   0.003389 435.935605 110.822  < 2e-16 ***
engines_DDG-Yandex     0.114274   0.003376 435.842691  33.851  < 2e-16 ***
engines_Google-Yahoo! -0.021268   0.003403 436.866813  -6.249 9.81e-10 ***
engines_Google-Yandex -0.046971   0.003387 435.645578 -13.868  < 2e-16 ***
engines_Yahoo!-Yandex  0.066890   0.003416 437.342607  19.579  < 2e-16 ***
---
Signif. codes:  0 '***' 0.001 '**' 0.01 '*' 0.05 '.' 0.1 ' ' 1
convergence code: 0
unable to evaluate scaled gradient
Model failed to converge: degenerate  Hessian with 1 negative eigenvalues

## Response:  RBO (p=0.8)

Linear mixed model fit by REML. t-tests use Satterthwaite's method ['lmerModLmerTest']
Formula: rbo_80 ~ engines_ + (1 | agent1) + (1 | agent2) + (1 | browser) +      (1 | machine_combination)
   Data: subdf

REML criterion at convergence: -27161.7

Scaled residuals:

```
    Min      1Q  Median      3Q     Max
-5.2171 -0.3499 -0.0027  0.4649  6.2392
```

Random effects:
```
 Groups              Name        Variance  Std.Dev.
 machine_combination (Intercept) 0.000e+00 0.000000
 agent2              (Intercept) 6.594e-04 0.025680
 agent1              (Intercept) 6.787e-04 0.026052
 browser             (Intercept) 8.563e-05 0.009254
 Residual                        1.443e-03 0.037992
Number of obs: 7677, groups:  machine_combination, 4074; agent2, 192; agent1, 192; browser, 2
```

Fixed effects:
```
                       Estimate Std. Error         df t value Pr(>|t|)
(Intercept)            0.096763   0.008215   1.811932  11.779 0.010108 *
engines_Baidu-DDG     -0.003761   0.005148 551.585037  -0.731 0.465387
engines_Baidu-Google  -0.089580   0.005185 551.493601 -17.278  < 2e-16 ***
engines_Baidu-Yahoo!   0.061160   0.005266 551.451972  11.615  < 2e-16 ***
engines_Baidu-Yandex  -0.095963   0.005227 552.560939 -18.360  < 2e-16 ***
engines_Bing-DDG      -0.016071   0.005149 552.023810  -3.121 0.001897 **
engines_Bing-Google    0.337638   0.005188 552.673115  65.075  < 2e-16 ***
engines_Bing-Yahoo!    0.067594   0.005265 551.553579  12.837  < 2e-16 ***
engines_Bing-Yandex   -0.094772   0.005224 551.551302 -18.140  < 2e-16 ***
engines_DDG-Google    -0.089291   0.006876 433.726362 -12.985  < 2e-16 ***
engines_DDG-Yahoo!     0.293810   0.006936 435.224124  42.358  < 2e-16 ***
engines_DDG-Yandex    -0.026629   0.006906 434.517786  -3.856 0.000133 ***
engines_Google-Yahoo! -0.088711   0.006962 435.585073 -12.742  < 2e-16 ***
engines_Google-Yandex -0.096472   0.006932 434.927955 -13.917  < 2e-16 ***
engines_Yahoo!-Yandex -0.053023   0.006993 436.719500  -7.583 2.05e-13 ***
---
Signif. codes:  0 '***' 0.001 '**' 0.01 '*' 0.05 '.' 0.1 ' ' 1
convergence code: 0
boundary (singular) fit: see ?isSingular
```

# D. Statistical Tests for "joe biden" query

## D.1 Comparison of Browsers for "joe biden" query

### Response: Jaccard

```
Linear mixed model fit by REML. t-tests use Satterthwaite's method ['lmerModLmerTest']
Formula: jaccard ~ browser * engine + (1 | agent1) + (1 | agent2) + (1 |      machine_combination)
   Data: subdf

REML criterion at convergence: -5942.5

Scaled residuals:
    Min      1Q  Median      3Q     Max
-5.3593 -0.2846  0.0002  0.2915 11.6410

Random effects:
```

```
 Groups             Name        Variance  Std.Dev.
 machine_combination (Intercept) 7.664e-05 0.008754
 agent2             (Intercept) 3.544e-03 0.059530
 agent1             (Intercept) 2.526e-03 0.050263
 Residual                       2.817e-03 0.053074
Number of obs: 2258, groups:  machine_combination, 1537; agent2, 192; agent1, 96

Fixed effects:
                             Estimate Std. Error         df t value Pr(>|t|)
(Intercept)                  0.882560   0.020280 264.879707  43.519  < 2e-16 ***
browserFirefox              -0.022267   0.022398 180.616863  -0.994 0.321484
engineBing                  -0.149840   0.029119 264.595654  -5.146 5.20e-07 ***
engineDDG                    0.105944   0.028141 261.908080   3.765 0.000206 ***
engineGoogle                 0.041277   0.028581 262.595999   1.444 0.149875
engineYahoo!                 0.117441   0.028630 264.108059   4.102 5.46e-05 ***
engineYandex                 0.015232   0.028597 263.081668   0.533 0.594727
browserFirefox:engineBing    0.037444   0.031667 181.019050   1.182 0.238589
browserFirefox:engineDDG     0.009032   0.030858 178.987662   0.293 0.770082
browserFirefox:engineGoogle  0.018143   0.031121 179.798153   0.583 0.560642
browserFirefox:engineYahoo!  0.018339   0.031630 180.273970   0.580 0.562771
browserFirefox:engineYandex  0.020996   0.031354 179.904044   0.670 0.503938
---
Signif. codes:  0 '***' 0.001 '**' 0.01 '*' 0.05 '.' 0.1 ' ' 1

Correlation of Fixed Effects:
            (Intr) brwsrF engnBn engDDG engnGg engnY! engnYn brwF:B bF:DDG brwF:G brF:Y!
browserFrfx -0.558
engineBing  -0.696  0.388
engineDDG   -0.721  0.402  0.502
engineGoogl -0.710  0.396  0.494  0.511
engineYaho! -0.708  0.395  0.493  0.510  0.503
engineYandx -0.709  0.396  0.494  0.511  0.503  0.502
brwsrFrfx:B  0.395 -0.707 -0.566 -0.284 -0.280 -0.279 -0.280
brwsrFr:DDG  0.405 -0.726 -0.282 -0.559 -0.287 -0.287 -0.287  0.513
brwsrFrfx:G  0.401 -0.720 -0.280 -0.289 -0.563 -0.284 -0.285  0.509  0.522
brwsrFrf:Y!  0.395 -0.708 -0.275 -0.285 -0.280 -0.556 -0.280  0.501  0.514  0.510
brwsrFrfx:Y  0.398 -0.714 -0.278 -0.287 -0.283 -0.282 -0.560  0.505  0.519  0.514  0.506
```

## Response: Jaccard Top-10

```
Linear mixed model fit by REML. t-tests use Satterthwaite's method ['lmerModLmerTest']
Formula: jacctop10 ~ browser * engine + (1 | agent1) + (1 | agent2) +      (1 | machine_combination)
   Data: subdf

REML criterion at convergence: -3565.6

Scaled residuals:
    Min      1Q  Median      3Q     Max
-5.8152 -0.4207  0.0000  0.3685  9.1425

Random effects:
 Groups              Name        Variance  Std.Dev.
 machine_combination (Intercept) 0.0005793 0.02407
 agent2              (Intercept) 0.0046714 0.06835
 agent1              (Intercept) 0.0138438 0.11766
 Residual                        0.0080593 0.08977
Number of obs: 2258, groups:  machine_combination, 1537; agent2, 192; agent1, 96
```

```
Fixed effects:
                             Estimate Std. Error         df t value Pr(>|t|)
(Intercept)                 8.657e-01  3.531e-02  1.636e+02  24.521  < 2e-16 ***
browserFirefox             -1.884e-02  2.700e-02  1.740e+02  -0.698  0.48617
engineBing                  6.676e-03  5.073e-02  1.635e+02   0.132  0.89547
engineDDG                  -3.006e-01  4.905e-02  1.619e+02  -6.129 6.51e-09 ***
engineGoogle                9.911e-02  4.981e-02  1.624e+02   1.990  0.04832 *
engineYahoo!                1.343e-01  4.989e-02  1.632e+02   2.691  0.00786 **
engineYandex                6.009e-02  4.984e-02  1.627e+02   1.206  0.22966
browserFirefox:engineBing   3.133e-03  3.817e-02  1.749e+02   0.082  0.93467
browserFirefox:engineDDG    1.104e-01  3.712e-02  1.720e+02   2.974  0.00336 **
browserFirefox:engineGoogle 3.695e-02  3.748e-02  1.736e+02   0.986  0.32565
browserFirefox:engineYahoo! 1.885e-02  3.812e-02  1.742e+02   0.494  0.62162
browserFirefox:engineYandex -6.289e-04 3.777e-02  1.737e+02  -0.017  0.98673
---
Signif. codes:  0 '***' 0.001 '**' 0.01 '*' 0.05 '.' 0.1 ' ' 1

Correlation of Fixed Effects:
            (Intr) brwsrF engnBn engDDG engnGg engnY! engnYn brwF:B bF:DDG brwF:G brF:Y!
browserFrfx -0.400
engineBing  -0.696  0.278
engineDDG   -0.720  0.288  0.501
engineGoogl -0.709  0.284  0.493  0.510
engineYaho! -0.708  0.283  0.493  0.509  0.502
engineYandx -0.708  0.283  0.493  0.510  0.502  0.501
brwsrFrfx:B  0.283 -0.705 -0.406 -0.204 -0.201 -0.200 -0.200
brwsrFr:DDG  0.291 -0.727 -0.202 -0.399 -0.206 -0.206 -0.206  0.513
brwsrFrfx:G  0.288 -0.720 -0.201 -0.207 -0.402 -0.204 -0.204  0.508  0.524
brwsrFrf:Y!  0.283 -0.708 -0.197 -0.204 -0.201 -0.399 -0.201  0.499  0.515  0.512
brwsrFrfx:Y  0.286 -0.715 -0.199 -0.206 -0.203 -0.202 -0.400  0.504  0.521  0.515  0.506
```

## Response:  RBO (p=0.95)

```
Linear mixed model fit by REML. t-tests use Satterthwaite's method ['lmerModLmerTest']
Formula: rbo_95 ~ browser * engine + (1 | agent1) + (1 | agent2) + (1 |      machine_combination)
   Data: subdf

REML criterion at convergence: -6401

Scaled residuals:
    Min      1Q  Median      3Q     Max
-5.9239 -0.3067 -0.0028  0.2066  7.3749

Random effects:
 Groups              Name        Variance Std.Dev.
 machine_combination (Intercept) 0.001332 0.03650
 agent2              (Intercept) 0.001048 0.03237
 agent1              (Intercept) 0.005200 0.07211
 Residual                        0.001350 0.03674
Number of obs: 2258, groups:  machine_combination, 1537; agent2, 192; agent1, 96

Fixed effects:
                  Estimate  Std. Error         df t value Pr(>|t|)
(Intercept)       0.816990    0.020434 138.374880  39.982  < 2e-16 ***
browserFirefox   -0.010696    0.013173 181.164801  -0.812  0.41788
engineBing       -0.030281    0.029369 138.335489  -1.031  0.30432
engineDDG        -0.138748    0.028400 137.145854  -4.885 2.84e-06 ***
engineGoogle      0.066865    0.028845 137.486227   2.318  0.02192 *
```

```
engineYahoo!                  -0.029367  0.028880 138.106590  -1.017  0.31100
engineYandex                   0.050720  0.028857 137.695743   1.758  0.08103 .
browserFirefox:engineBing      0.006800  0.018391 173.754399   0.370  0.71202
browserFirefox:engineDDG       0.053409  0.018099 178.891318   2.951  0.00359 **
browserFirefox:engineGoogle    0.013307  0.018289 181.152457   0.728  0.46781
browserFirefox:engineYahoo!    0.008767  0.018600 181.505112   0.471  0.63797
browserFirefox:engineYandex    0.011409  0.018427 181.152354   0.619  0.53660
---
Signif. codes:  0 '***' 0.001 '**' 0.01 '*' 0.05 '.' 0.1 ' ' 1

Correlation of Fixed Effects:
            (Intr) brwsrF engnBn engDDG engnGg engnY! engnYn brwF:B bF:DDG brwF:G brF:Y!
browserFrfx -0.344
engineBing  -0.696  0.239
engineDDG   -0.719  0.247  0.501
engineGoogl -0.708  0.243  0.493  0.510
engineYaho! -0.708  0.243  0.492  0.509  0.501
engineYandx -0.708  0.243  0.493  0.509  0.502  0.501
brwsrFrfx:B  0.246 -0.695 -0.354 -0.177 -0.175 -0.174 -0.175
brwsrFr:DDG  0.250 -0.728 -0.174 -0.342 -0.177 -0.177 -0.177  0.506
brwsrFrfx:G  0.248 -0.720 -0.172 -0.178 -0.345 -0.175 -0.175  0.501  0.524
brwsrFrf:Y!  0.243 -0.708 -0.169 -0.175 -0.173 -0.343 -0.172  0.492  0.515  0.525
brwsrFrfx:Y  0.246 -0.715 -0.171 -0.177 -0.174 -0.174 -0.344  0.497  0.534  0.515  0.506
convergence code: 0
Model failed to converge with max|grad| = 0.00207129 (tol = 0.002, component 1)
```

## Response: RBO (p=0.8)

```
Linear mixed model fit by REML. t-tests use Satterthwaite's method ['lmerModLmerTest']
Formula: rbo_80 ~ browser * engine + (1 | agent1) + (1 | agent2) + (1 |      machine_combination)
   Data: subdf

REML criterion at convergence: -3731.5

Scaled residuals:
    Min      1Q  Median      3Q     Max
-5.8694 -0.3108  0.0043  0.2445  8.4232

Random effects:
 Groups              Name        Variance  Std.Dev.
 machine_combination (Intercept) 0.0000000 0.00000
 agent2              (Intercept) 0.0002155 0.01468
 agent1              (Intercept) 0.0086548 0.09303
 Residual                        0.0093675 0.09679
Number of obs: 2258, groups:  machine_combination, 1537; agent2, 192; agent1, 96

Fixed effects:
                            Estimate Std. Error         df t value Pr(>|t|)
(Intercept)                9.503e-01  2.525e-02  1.138e+02  37.641  < 2e-16 ***
browserFirefox            -5.239e-03  1.224e-02  1.969e+02  -0.428  0.66908
engineBing                 7.142e-03  3.635e-02  1.145e+02   0.196  0.84457
engineDDG                 -4.567e-01  3.508e-02  1.126e+02 -13.019  < 2e-16 ***
engineGoogle               2.936e-02  3.567e-02  1.134e+02   0.823  0.41212
engineYahoo!               1.665e-02  3.570e-02  1.138e+02   0.466  0.64192
engineYandex               6.755e-06  3.568e-02  1.136e+02   0.000  0.99985
browserFirefox:engineBing  3.581e-03  1.758e-02  2.128e+02   0.204  0.83879
browserFirefox:engineDDG   6.292e-02  1.670e-02  1.911e+02   3.768  0.00022 ***
```

```
browserFirefox:engineGoogle  1.421e-03  1.705e-02  2.022e+02   0.083  0.93367
browserFirefox:engineYahoo!  4.672e-03  1.731e-02  1.977e+02   0.270  0.78755
browserFirefox:engineYandex  5.939e-03  1.718e-02  2.009e+02   0.346  0.72985
---
Signif. codes:  0 '***' 0.001 '**' 0.01 '*' 0.05 '.' 0.1 ' ' 1

Correlation of Fixed Effects:
            (Intr) brwsrF engnBn engDDG engnGg engnY! engnYn brwF:B bF:DDG brwF:G brF:Y!
browserFrfx -0.312
engineBing  -0.695  0.217
engineDDG   -0.720  0.225  0.500
engineGoogl -0.708  0.221  0.492  0.509
engineYaho! -0.707  0.221  0.491  0.509  0.501
engineYandx -0.707  0.221  0.491  0.509  0.501  0.500
brwsrFrfx:B  0.217 -0.696 -0.318 -0.156 -0.154 -0.154 -0.154
brwsrFr:DDG  0.229 -0.733 -0.159 -0.308 -0.162 -0.162 -0.162  0.510
brwsrFrfx:G  0.224 -0.718 -0.156 -0.161 -0.313 -0.158 -0.158  0.500  0.526
brwsrFrf:Y!  0.221 -0.707 -0.153 -0.159 -0.156 -0.312 -0.156  0.492  0.518  0.508
brwsrFrfx:Y  0.222 -0.713 -0.154 -0.160 -0.157 -0.157 -0.313  0.496  0.522  0.512  0.504
convergence code: 0
boundary (singular) fit: see ?isSingular
```

## D.2 Comparison of Search Engines for "joe biden" query

### Response: Jaccard

```
Linear mixed model fit by REML. t-tests use Satterthwaite's method ['lmerModLmerTest']
Formula: jaccard ~ engines_ + (1 | agent1) + (1 | agent2) + (1 | browser) +      (1 | machine_combination)
   Data: subdf

REML criterion at convergence: -44799.8

Scaled residuals:
    Min      1Q  Median      3Q     Max
-7.8027 -0.2687  0.0181  0.2931  5.1795

Random effects:
 Groups              Name        Variance  Std.Dev.
 machine_combination (Intercept) 6.004e-06 0.002450
 agent2              (Intercept) 4.396e-05 0.006630
 agent1              (Intercept) 4.003e-05 0.006327
 browser             (Intercept) 0.000e+00 0.000000
 Residual                        1.419e-04 0.011910
Number of obs: 7677, groups:  machine_combination, 4074; agent2, 192; agent1, 192; browser, 2

Fixed effects:
                     Estimate Std. Error         df t value Pr(>|t|)
(Intercept)         2.523e-02  1.296e-03  4.815e+02   19.46   <2e-16 ***
engines_Baidu-DDG   1.864e-02  1.382e-03  6.759e+02   13.49   <2e-16 ***
engines_Baidu-Google 2.067e-02 1.389e-03  6.721e+02   14.87   <2e-16 ***
engines_Baidu-Yahoo! 4.460e-02 1.413e-03  6.752e+02   31.57   <2e-16 ***
engines_Baidu-Yandex 3.771e-02 1.401e-03  6.742e+02   26.91   <2e-16 ***
engines_Bing-DDG    6.695e-02  1.381e-03  6.741e+02   48.47   <2e-16 ***
engines_Bing-Google 7.667e-02  1.390e-03  6.721e+02   55.16   <2e-16 ***
engines_Bing-Yahoo! 6.347e-02  1.413e-03  6.740e+02   44.93   <2e-16 ***
```

```
engines_Bing-Yandex      2.847e-02  1.400e-03 6.715e+02   20.34   <2e-16 ***
engines_DDG-Google       8.808e-02  1.789e-03 4.714e+02   49.25   <2e-16 ***
engines_DDG-Yahoo!       5.796e-01  1.805e-03 4.739e+02  321.07   <2e-16 ***
engines_DDG-Yandex       8.781e-02  1.799e-03 4.747e+02   48.80   <2e-16 ***
engines_Google-Yahoo!    1.150e-01  1.816e-03 4.779e+02   63.30   <2e-16 ***
engines_Google-Yandex    1.149e-01  1.804e-03 4.737e+02   63.67   <2e-16 ***
engines_Yahoo!-Yandex    1.042e-01  1.820e-03 4.763e+02   57.25   <2e-16 ***
---
Signif. codes:  0 '***' 0.001 '**' 0.01 '*' 0.05 '.' 0.1 ' ' 1
convergence code: 0
boundary (singular) fit: see ?isSingular
```

## Response: Jaccard Top-10

```
Linear mixed model fit by REML. t-tests use Satterthwaite's method ['lmerModLmerTest']
Formula: jacctop10 ~ engines_ + (1 | agent1) + (1 | agent2) + (1 | browser) +      (1 | machine_combination)
   Data: subdf

REML criterion at convergence: -27485.4

Scaled residuals:
    Min      1Q  Median      3Q     Max
-3.7483 -0.3324  0.0608  0.2375 11.4652

Random effects:
 Groups              Name        Variance  Std.Dev.
 machine_combination (Intercept) 0.000e+00 0.000000
 agent2              (Intercept) 7.228e-04 0.026885
 agent1              (Intercept) 4.372e-04 0.020910
 browser             (Intercept) 6.581e-05 0.008113
 Residual                        1.392e-03 0.037316
Number of obs: 7677, groups:  machine_combination, 4074; agent2, 192; agent1, 192; browser, 2

Fixed effects:
                         Estimate Std. Error        df t value Pr(>|t|)
(Intercept)             5.301e-02  7.339e-03 1.898e+00   7.224   0.0214 *
engines_Baidu-DDG       1.607e-03  4.752e-03 5.523e+02   0.338   0.7353
engines_Baidu-Google   -6.313e-05  4.779e-03 5.491e+02  -0.013   0.9895
engines_Baidu-Yahoo!    9.597e-02  4.860e-03 5.525e+02  19.748   <2e-16 ***
engines_Baidu-Yandex    5.846e-02  4.818e-03 5.500e+02  12.133   <2e-16 ***
engines_Bing-DDG        5.939e-02  4.747e-03 5.488e+02  12.510   <2e-16 ***
engines_Bing-Google     1.258e-01  4.779e-03 5.474e+02  26.328   <2e-16 ***
engines_Bing-Yahoo!     1.495e-01  4.856e-03 5.488e+02  30.799   <2e-16 ***
engines_Bing-Yandex     1.251e-01  4.813e-03 5.466e+02  26.001   <2e-16 ***
engines_DDG-Google      3.537e-04  6.282e-03 4.162e+02   0.056   0.9551
engines_DDG-Yahoo!      1.960e-01  6.337e-03 4.179e+02  30.928   <2e-16 ***
engines_DDG-Yandex      7.387e-02  6.308e-03 4.170e+02  11.710   <2e-16 ***
engines_Google-Yahoo!   7.327e-02  6.364e-03 4.190e+02  11.514   <2e-16 ***
engines_Google-Yandex   1.620e-03  6.334e-03 4.178e+02   0.256   0.7983
engines_Yahoo!-Yandex   1.767e-01  6.390e-03 4.196e+02  27.659   <2e-16 ***
---
Signif. codes:  0 '***' 0.001 '**' 0.01 '*' 0.05 '.' 0.1 ' ' 1
convergence code: 0
boundary (singular) fit: see ?isSingular
```

## Response: RBO (p=0.95)

```
Linear mixed model fit by REML. t-tests use Satterthwaite's method ['lmerModLmerTest']
Formula: rbo_95 ~ engines_ + (1 | agent1) + (1 | agent2) + (1 | browser) +      (1 | machine_combination)
   Data: subdf

REML criterion at convergence: -38088.9

Scaled residuals:
    Min      1Q  Median      3Q     Max
-4.9635 -0.3656 -0.0118  0.3108  7.9731

Random effects:
 Groups              Name        Variance  Std.Dev.
 machine_combination (Intercept) 0.0000000 0.00000
 agent2              (Intercept) 0.0005276 0.02297
 agent1              (Intercept) 0.0004231 0.02057
 browser             (Intercept) 0.0001227 0.01108
 Residual                        0.0003286 0.01813
Number of obs: 7677, groups:  machine_combination, 4074; agent2, 192; agent1, 192; browser, 2

Fixed effects:
                      Estimate Std. Error        df t value Pr(>|t|)
(Intercept)           4.771e-02  8.791e-03 1.296e+00   5.427 0.075834 .
engines_Baidu-DDG     1.742e-02  3.979e-03 4.256e+02   4.377 1.51e-05 ***
engines_Baidu-Google  1.525e-02  4.006e-03 4.248e+02   3.808 0.000161 ***
engines_Baidu-Yahoo!  8.164e-02  4.070e-03 4.255e+02  20.059  < 2e-16 ***
engines_Baidu-Yandex  7.459e-02  4.037e-03 4.252e+02  18.475  < 2e-16 ***
engines_Bing-DDG      8.885e-02  3.978e-03 4.251e+02  22.333  < 2e-16 ***
engines_Bing-Google   2.750e-01  4.007e-03 4.247e+02  68.634  < 2e-16 ***
engines_Bing-Yahoo!   1.415e-01  4.069e-03 4.251e+02  34.783  < 2e-16 ***
engines_Bing-Yandex   1.465e-01  4.036e-03 4.246e+02  36.286  < 2e-16 ***
engines_DDG-Google    8.534e-02  5.521e-03 3.886e+02  15.458  < 2e-16 ***
engines_DDG-Yahoo!    3.426e-01  5.566e-03 3.891e+02  61.553  < 2e-16 ***
engines_DDG-Yandex    1.442e-01  5.542e-03 3.888e+02  26.013  < 2e-16 ***
engines_Google-Yahoo! 1.955e-01  5.586e-03 3.894e+02  34.993  < 2e-16 ***
engines_Google-Yandex 1.077e-01  5.563e-03 3.891e+02  19.352  < 2e-16 ***
engines_Yahoo!-Yandex 2.369e-01  5.608e-03 3.896e+02  42.239  < 2e-16 ***
---
Signif. codes:  0 '***' 0.001 '**' 0.01 '*' 0.05 '.' 0.1 ' ' 1
convergence code: 0
boundary (singular) fit: see ?isSingular
```

## Response: RBO (p=0.8)

```
Linear mixed model fit by REML. t-tests use Satterthwaite's method ['lmerModLmerTest']
Formula: rbo_80 ~ engines_ + (1 | agent1) + (1 | agent2) + (1 | browser) +      (1 | machine_combination)
   Data: subdf

REML criterion at convergence: -24428.7

Scaled residuals:
    Min      1Q  Median      3Q     Max
-5.5253 -0.3632 -0.0001  0.3235  5.8919

Random effects:
 Groups              Name        Variance  Std.Dev.
 machine_combination (Intercept) 0.0000000 0.00000
```

```
agent2              (Intercept) 0.0021570 0.04644
agent1              (Intercept) 0.0017845 0.04224
browser             (Intercept) 0.0006293 0.02508
Residual                        0.0019896 0.04461
Number of obs: 7677, groups:  machine_combination, 4074; agent2, 192; agent1, 192; browser, 2

Fixed effects:
                      Estimate Std. Error         df t value Pr(>|t|)
(Intercept)          2.971e-02  1.955e-02  1.301e+00   1.520   0.3269
engines_Baidu-DDG    8.604e-04  8.271e-03  4.532e+02   0.104   0.9172
engines_Baidu-Google 2.315e-03  8.325e-03  4.520e+02   0.278   0.7810
engines_Baidu-Yahoo! 3.985e-02  8.460e-03  4.531e+02   4.711 3.29e-06 ***
engines_Baidu-Yandex 1.977e-02  8.391e-03  4.527e+02   2.356   0.0189 *
engines_Bing-DDG     8.891e-02  8.269e-03  4.526e+02  10.753  < 2e-16 ***
engines_Bing-Google  5.622e-01  8.327e-03  4.520e+02  67.525  < 2e-16 ***
engines_Bing-Yahoo!  2.169e-01  8.457e-03  4.526e+02  25.644  < 2e-16 ***
engines_Bing-Yandex  2.246e-01  8.388e-03  4.518e+02  26.772  < 2e-16 ***
engines_DDG-Google   6.660e-02  1.137e-02  3.987e+02   5.859 9.75e-09 ***
engines_DDG-Yahoo!   2.297e-01  1.146e-02  3.994e+02  20.042  < 2e-16 ***
engines_DDG-Yandex   1.446e-01  1.141e-02  3.990e+02  12.666  < 2e-16 ***
engines_Google-Yahoo! 2.875e-01 1.151e-02  3.998e+02  24.989  < 2e-16 ***
engines_Google-Yandex 9.405e-02 1.146e-02  3.993e+02   8.209 3.11e-15 ***
engines_Yahoo!-Yandex 4.181e-01 1.155e-02  4.001e+02  36.201  < 2e-16 ***
---
Signif. codes:  0 '***' 0.001 '**' 0.01 '*' 0.05 '.' 0.1 ' ' 1
convergence code: 0
boundary (singular) fit: see ?isSingular
```

# E. Comparison of politician-like queries withing search engines

## Response: Jaccard

```
Linear mixed model fit by REML. t-tests use Satterthwaite's method ['lmerModLmerTest']
Formula: jaccard ~ politician + (1 | engines_) + (1 | browser) + (1 |
    agent1) + (1 | agent2) + (1 | agent_combination) + (1 | machine_combination)
   Data: subdf

REML criterion at convergence: -20453.6

Scaled residuals:
    Min      1Q  Median      3Q     Max
-4.9468 -0.4452  0.0552  0.5394  5.9005

Random effects:
 Groups              Name        Variance  Std.Dev.
 agent_combination   (Intercept) 0.0005174 0.02275
 machine_combination (Intercept) 0.0000000 0.00000
 agent2              (Intercept) 0.0016921 0.04114
 agent1              (Intercept) 0.0016931 0.04115
 engines_            (Intercept) 0.0051241 0.07158
 browser             (Intercept) 0.0000000 0.00000
 Residual                        0.0044819 0.06695
```

```
Number of obs: 8658, groups:  agent_combination, 1443; machine_combination, 768; agent2, 192; agent1, 192; engines_, 6;
browser, 2

Fixed effects:
                      Estimate Std. Error         df t value Pr(>|t|)
(Intercept)          9.073e-01  2.956e-02  5.042e+00  30.698 6.27e-07 ***
politicianDonald Trump 8.521e-03  1.762e-03  7.051e+03   4.835 1.36e-06 ***
politicianJoe Biden   -8.172e-03  1.762e-03  7.051e+03  -4.637 3.60e-06 ***
---
Signif. codes:  0 '***' 0.001 '**' 0.01 '*' 0.05 '.' 0.1 ' ' 1

Correlation of Fixed Effects:
            (Intr) pltcDT
pltcnDnldTr -0.030
politcnJBdn -0.030  0.500
convergence code: 0
boundary (singular) fit: see ?isSingular
```

## Response: Jaccard Top-10

```
Linear mixed model fit by REML. t-tests use Satterthwaite's method ['lmerModLmerTest']
Formula: jacctop10 ~ politician + (1 | engines_) + (1 | browser) + (1 |
    agent1) + (1 | agent2) + (1 | agent_combination) + (1 | machine_combination)
   Data: subdf

REML criterion at convergence: -9530.5

Scaled residuals:
    Min      1Q  Median      3Q     Max
-4.8028 -0.5167  0.0872  0.5867  4.8079

Random effects:
 Groups              Name        Variance Std.Dev.
 agent_combination   (Intercept) 0.004049 0.06363
 machine_combination (Intercept) 0.000000 0.00000
 agent2              (Intercept) 0.004252 0.06521
 agent1              (Intercept) 0.004252 0.06520
 engines_            (Intercept) 0.011701 0.10817
 browser             (Intercept) 0.002483 0.04983
 Residual                        0.015008 0.12251
Number of obs: 8658, groups:  agent_combination, 1443; machine_combination, 768; agent2, 192; agent1, 192; engines_, 6;
browser, 2

Fixed effects:
                      Estimate Std. Error         df t value Pr(>|t|)
(Intercept)          8.421e-01  5.696e-02  4.274e+00   14.79 7.85e-05 ***
politicianDonald Trump -1.580e-02  3.225e-03  7.056e+03   -4.90 9.82e-07 ***
politicianJoe Biden    4.239e-02  3.225e-03  7.056e+03   13.14  < 2e-16 ***
---
Signif. codes:  0 '***' 0.001 '**' 0.01 '*' 0.05 '.' 0.1 ' ' 1

Correlation of Fixed Effects:
            (Intr) pltcDT
pltcnDnldTr -0.028
politcnJBdn -0.028  0.500
convergence code: 0
boundary (singular) fit: see ?isSingular
```

## Response: RBO (p=0.95)

```
Linear mixed model fit by REML. t-tests use Satterthwaite's method ['lmerModLmerTest']
Formula: rbo_95 ~ politician + (1 | engines_) + (1 | browser) + (1 | agent1) +
    (1 | agent2) + (1 | agent_combination) + (1 | machine_combination)
   Data: subdf

REML criterion at convergence: -18437.2

Scaled residuals:
    Min      1Q  Median      3Q     Max
-5.4949 -0.4546  0.0551  0.5333  5.6292

Random effects:
 Groups              Name        Variance  Std.Dev.
 agent_combination   (Intercept) 0.0019422 0.04407
 machine_combination (Intercept) 0.0000000 0.00000
 agent2              (Intercept) 0.0014980 0.03870
 agent1              (Intercept) 0.0014976 0.03870
 engines_            (Intercept) 0.0017116 0.04137
 browser             (Intercept) 0.0008291 0.02879
 Residual                        0.0051943 0.07207
Number of obs: 8658, groups:  agent_combination, 1443; machine_combination, 768; agent2, 192; agent1, 192; engines_, 6; browser, 2

Fixed effects:
                       Estimate Std. Error        df t value Pr(>|t|)
(Intercept)           7.948e-01  2.681e-02 2.532e+00   29.65 0.000276 ***
politicianDonald Trump -3.726e-02 1.897e-03 7.056e+03  -19.64  < 2e-16 ***
politicianJoe Biden    2.012e-02  1.897e-03 7.056e+03   10.60  < 2e-16 ***
---
Signif. codes:  0 '***' 0.001 '**' 0.01 '*' 0.05 '.' 0.1 ' ' 1

Correlation of Fixed Effects:
            (Intr) pltcDT
pltcnDnldTr -0.035
politcnJBdn -0.035  0.500
convergence code: 0
boundary (singular) fit: see ?isSingular
```

## Response: RBO (p=0.8)

```
Linear mixed model fit by REML. t-tests use Satterthwaite's method ['lmerModLmerTest']
Formula: rbo_80 ~ politician + (1 | engines_) + (1 | browser) + (1 | agent1) +
    (1 | agent2) + (1 | agent_combination) + (1 | machine_combination)
   Data: subdf

REML criterion at convergence: -9310.2

Scaled residuals:
    Min      1Q  Median      3Q     Max
-5.1111 -0.4612  0.0477  0.5787  5.0111

Random effects:
 Groups              Name        Variance  Std.Dev.
 agent_combination   (Intercept) 1.261e-02 1.123e-01
```

```
machine_combination (Intercept) 1.819e-05 4.265e-03
agent2              (Intercept) 9.341e-10 3.056e-05
agent1              (Intercept) 0.000e+00 0.000e+00
engines_            (Intercept) 1.732e-02 1.316e-01
browser             (Intercept) 1.578e-03 3.973e-02
Residual                        1.465e-02 1.210e-01
```
Number of obs: 8658, groups:  agent_combination, 1443; machine_combination, 768; agent2, 192; agent1, 192; engines_, 6; browser, 2

Fixed effects:
```
                        Estimate Std. Error         df t value Pr(>|t|)
(Intercept)            8.511e-01  6.075e-02  5.903e+00   14.01 9.4e-06 ***
politicianDonald Trump -6.360e-02  3.186e-03  7.213e+03  -19.96  < 2e-16 ***
politicianJoe Biden     5.165e-02  3.186e-03  7.213e+03   16.21  < 2e-16 ***
---
Signif. codes:  0 '***' 0.001 '**' 0.01 '*' 0.05 '.' 0.1 ' ' 1
```

Correlation of Fixed Effects:
```
            (Intr) pltcDT
pltcnDnldTr -0.026
politcnJBdn -0.026  0.500
convergence code: 0
boundary (singular) fit: see ?isSingular
```

# F. Comparison of politician-like queries controlling by all factors

## Response: Jaccard

Linear mixed model fit by REML. t-tests use Satterthwaite's method ['lmerModLmerTest']
Formula: jaccard ~ politician + (1 | engines_) + (1 | browser) + (1 |
    agent1) + (1 | agent2) + (1 | agent_combination) + (1 | machine_combination)
   Data: subdf

REML criterion at convergence: -20453.6

Scaled residuals:
```
    Min      1Q  Median      3Q     Max
-4.9468 -0.4452  0.0552  0.5394  5.9005
```

Random effects:
```
 Groups              Name        Variance  Std.Dev.
 agent_combination   (Intercept) 0.0005174 0.02275
 machine_combination (Intercept) 0.0000000 0.00000
 agent2              (Intercept) 0.0016921 0.04114
 agent1              (Intercept) 0.0016931 0.04115
 engines_            (Intercept) 0.0051241 0.07158
 browser             (Intercept) 0.0000000 0.00000
 Residual                        0.0044819 0.06695
```
Number of obs: 8658, groups:  agent_combination, 1443; machine_combination, 768; agent2, 192; agent1, 192; engines_, 6; browser, 2

Fixed effects:

```
                    Estimate Std. Error         df t value Pr(>|t|)
(Intercept)            9.073e-01  2.956e-02  5.042e+00   30.698 6.27e-07 ***
politicianDonald Trump 8.521e-03  1.762e-03  7.051e+03    4.835 1.36e-06 ***
politicianJoe Biden   -8.172e-03  1.762e-03  7.051e+03   -4.637 3.60e-06 ***
---
Signif. codes:  0 '***' 0.001 '**' 0.01 '*' 0.05 '.' 0.1 ' ' 1

Correlation of Fixed Effects:
            (Intr) pltcDT
pltcnDnldTr -0.030
politcnJBdn -0.030  0.500
convergence code: 0
boundary (singular) fit: see ?isSingular
```

## Response: Jaccard Top-10

```
Linear mixed model fit by REML. t-tests use Satterthwaite's method ['lmerModLmerTest']
Formula: jacctop10 ~ politician + (1 | engines_) + (1 | browser) + (1 |
    agent1) + (1 | agent2) + (1 | agent_combination) + (1 | machine_combination)
   Data: subdf

REML criterion at convergence: -9530.5

Scaled residuals:
    Min      1Q  Median      3Q     Max
-4.8028 -0.5167  0.0872  0.5867  4.8079

Random effects:
 Groups              Name        Variance Std.Dev.
 agent_combination   (Intercept) 0.004049 0.06363
 machine_combination (Intercept) 0.000000 0.00000
 agent2              (Intercept) 0.004252 0.06521
 agent1              (Intercept) 0.004252 0.06520
 engines_            (Intercept) 0.011701 0.10817
 browser             (Intercept) 0.002483 0.04983
 Residual                        0.015008 0.12251
Number of obs: 8658, groups:  agent_combination, 1443; machine_combination, 768; agent2, 192; agent1, 192; engines_, 6; browser, 2

Fixed effects:
                       Estimate Std. Error         df t value Pr(>|t|)
(Intercept)            8.421e-01  5.696e-02  4.274e+00   14.79 7.85e-05 ***
politicianDonald Trump -1.580e-02  3.225e-03  7.056e+03   -4.90 9.82e-07 ***
politicianJoe Biden     4.239e-02  3.225e-03  7.056e+03   13.14  < 2e-16 ***
---
Signif. codes:  0 '***' 0.001 '**' 0.01 '*' 0.05 '.' 0.1 ' ' 1

Correlation of Fixed Effects:
            (Intr) pltcDT
pltcnDnldTr -0.028
politcnJBdn -0.028  0.500
convergence code: 0
boundary (singular) fit: see ?isSingular
```

## Response: RBO (p=0.95)

Linear mixed model fit by REML. t-tests use Satterthwaite's method ['lmerModLmerTest']

```
Formula: rbo_95 ~ politician + (1 | engines_) + (1 | browser) + (1 | agent1) +
    (1 | agent2) + (1 | agent_combination) + (1 | machine_combination)
   Data: subdf

REML criterion at convergence: -18437.2

Scaled residuals:
    Min     1Q  Median     3Q    Max
-5.4949 -0.4546  0.0551  0.5333  5.6292

Random effects:
 Groups              Name        Variance  Std.Dev.
 agent_combination   (Intercept) 0.0019422 0.04407
 machine_combination (Intercept) 0.0000000 0.00000
 agent2              (Intercept) 0.0014980 0.03870
 agent1              (Intercept) 0.0014976 0.03870
 engines_            (Intercept) 0.0017116 0.04137
 browser             (Intercept) 0.0008291 0.02879
 Residual                        0.0051943 0.07207
Number of obs: 8658, groups:  agent_combination, 1443; machine_combination, 768; agent2, 192; agent1, 192; engines_, 6; browser, 2

Fixed effects:
                        Estimate Std. Error        df t value Pr(>|t|)
(Intercept)            7.948e-01  2.681e-02 2.532e+00   29.65 0.000276 ***
politicianDonald Trump -3.726e-02  1.897e-03 7.056e+03  -19.64  < 2e-16 ***
politicianJoe Biden     2.012e-02  1.897e-03 7.056e+03   10.60  < 2e-16 ***
---
Signif. codes:  0 '***' 0.001 '**' 0.01 '*' 0.05 '.' 0.1 ' ' 1

Correlation of Fixed Effects:
            (Intr) pltcDT
pltcnDnldTr -0.035
politcnJBdn -0.035  0.500
convergence code: 0
boundary (singular) fit: see ?isSingular
```

## Response: RBO (p=0.8)

```
Linear mixed model fit by REML. t-tests use Satterthwaite's method ['lmerModLmerTest']
Formula: rbo_80 ~ politician + (1 | engines_) + (1 | browser) + (1 | agent1) +
    (1 | agent2) + (1 | agent_combination) + (1 | machine_combination)
   Data: subdf

REML criterion at convergence: -9310.2

Scaled residuals:
    Min     1Q  Median     3Q    Max
-5.1111 -0.4612  0.0477  0.5787  5.0111

Random effects:
 Groups              Name        Variance  Std.Dev.
 agent_combination   (Intercept) 1.261e-02 1.123e-01
 machine_combination (Intercept) 1.819e-05 4.265e-03
 agent2              (Intercept) 9.341e-10 3.056e-05
 agent1              (Intercept) 0.000e+00 0.000e+00
 engines_            (Intercept) 1.732e-02 1.316e-01
```

```
 browser           (Intercept) 1.578e-03 3.973e-02
 Residual                      1.465e-02 1.210e-01
Number of obs: 8658, groups:  agent_combination, 1443; machine_combination, 768; agent2, 192; agent1, 192; engines_, 6; browser, 2

Fixed effects:
                         Estimate Std. Error         df t value Pr(>|t|)
(Intercept)             8.511e-01  6.075e-02  5.903e+00   14.01 9.4e-06 ***
politicianDonald Trump -6.360e-02  3.186e-03  7.213e+03  -19.96  < 2e-16 ***
politicianJoe Biden     5.165e-02  3.186e-03  7.213e+03   16.21  < 2e-16 ***
---
Signif. codes:  0 '***' 0.001 '**' 0.01 '*' 0.05 '.' 0.1 ' ' 1

Correlation of Fixed Effects:
            (Intr) pltcDT
pltcnDnldTr -0.026
politcnJBdn -0.026  0.500
convergence code: 0
boundary (singular) fit: see ?isSingular
```